\documentclass[final,3p,times]{elsarticle}

\usepackage{amsmath}
\usepackage{amssymb}
\usepackage{amsthm}
\graphicspath{{fig/}}
\usepackage{subcaption}
\usepackage{esint}
\usepackage{float}
\usepackage{multirow}
\usepackage{rotating}
\usepackage{booktabs}
\usepackage{xcolor}
\usepackage{tabularx}
\usepackage{lineno}
\usepackage{algorithm}
\usepackage{algpseudocode}
\usepackage[normalem]{ulem}
\usepackage{makecell}

\journal{Elsevier}

\begin{document}

\begin{frontmatter}


\title{Multi-Resolution Training-Enhanced Kolmogorov--Arnold Networks for Physics-Informed Learning of PDEs with Multi-Scale Features}

\affiliation[labelTJ]{organization={College of Civil Engineering, Tongji University},
	city={Shanghai},
	postcode={200092}, 
	country={China}}

\affiliation[labelSNU]{organization={Department of Mathematics, Shanghai Normal University},
	city={Shanghai},
	postcode={200234}, 
	country={China}}

\author[labelTJ]{Yu-Sen Yang}
\author[labelSNU]{Ling Guo}
\author[labelTJ]{Xiaodan Ren\corref{cor1}}
\ead{rxdtj@tongji.edu.cn}

\cortext[cor1]{Corresponding author.}

\begin{abstract}
PDE problems with multi-scale features present significant challenges in scientific computing. While conventional deep learning methods based on multi-layer perceptrons exhibit spectral bias in resolving multi-scale features, physics-informed Kolmogorov--Arnold networks mitigate this issue through their architectural design and show advantages in representing multi-scale physical fields. On the other hand, insights from information bottleneck theory suggest that high-resolution sampling is important for capturing multi-scale behavior, but this requirement often leads to increased computational cost. To address this issue, we propose a simple yet effective multi-resolution training-enhanced Kolmogorov--Arnold network framework, which trains the model sequentially or alternately at different resolutions. The proposed framework is validated on various forward and inverse PDE problems with multi-scale features. Numerical results indicate that this training strategy effectively reduces computational cost without sacrificing accuracy, thereby enabling the efficient solution of complex multi-scale PDE problems in both forward and inverse settings.
\end{abstract}
 

\end{frontmatter}

\section{Introduction}
\label{sec:Introduction}
Artificial intelligence has become an important tool in engineering and scientific computing, particularly for data-driven prediction and surrogate modelling of complex physical systems~\cite{singhDeepMachineLearning2024, deshpandePerformanceLifeAnalysis2025}.	Within this broader development, AI for PDEs has emerged as an important framework for accelerating the solution and inverse analysis of PDE-governed systems~\cite{wangArtificialIntelligencePartial}.
Its core idea is to incorporate governing equations, boundary conditions, and physical constraints into neural network training, thereby reducing the reliance on high-quality labeled data and improving the physical consistency of PDE solutions.
Representative works include the Deep Ritz Method~\cite{E2018}, Physics-Informed Neural Networks (PINNs)~\cite{raissiPhysicsinformedNeuralNetworks2019,Karniadakis2021}, and the Deep Energy Method~\cite{Samaniego2020,wangCENNConservativeEnergy2022,wangDCEMDeepComplementary2024,le-ducNormalizedEnergybasedPhysicsinformed2026}, among others.

Although PINNs~\cite{raissiPhysicsinformedNeuralNetworks2019,Karniadakis2021} have been successfully formulated for various forward and inverse PDE problems, their efficient training remains challenging for PDEs whose solutions or parameter fields contain high-frequency or multi-scale features~\cite{Rahaman2019, Wang2021b}. Such features may manifest as high-frequency oscillations, complex spatial patterns, or sharp material-property variations, and can pose substantial challenges to the training of PINNs. The main difficulties of PINNs for such problems can be broadly categorized into two aspects.
First, neural networks often struggle to effectively capture multi-scale components in physical fields or signals.
Second, multi-scale physical systems require a significantly large number of sampling points for training, resulting in substantial computational burdens.

The analysis of~\cite{Rahaman2019,Wang2021b} revealed that Multi-Layer Perceptrons (MLPs) exhibit significantly slower convergence for high-frequency components, highlighting the spectral bias pathology.
To mitigate this limitation, several modifications to standard MLP architectures have been developed, including adaptive activation functions, input scaling techniques, and Fourier feature mappings~\cite{Wang2021b,Jagtap2020,  huangFrequencyadaptiveMultiscaleDeep2025}. More broadly, Xu et al.~\cite{xuUnderstandingOvercomingSpectral2025a} systematically classified these strategies into frequency-shifting and frequency-scaling approaches.
As an alternative to MLPs, recent studies have revisited Kolmogorov--Arnold Networks (KANs)~\cite{KAN2024a,KAN2024b}, which use learnable univariate functions instead of fixed activation functions. Several KAN variants have been developed by using different parameterizations of the learnable univariate activation functions~\cite{FAIRKAN2024,RBFKAN,WaveletKAN,ChebyKAN}.
As reported in~\cite{wangKolmogorovArnoldInformed2025}, Physics-Informed Kolmogorov--Arnold Networks (PIKANs) exhibit reduced spectral bias compared with MLPs and offer certain advantages in modelling multi-scale problems.

Although improved MLP architectures and KANs have enhanced the representation of multi-scale physical fields, accurately resolving multi-scale features in physics-informed training still requires dense residual sampling, which limits their practical applicability.
Adaptive sampling algorithms~\cite{Wu2023} have been developed to improve sampling efficiency by allocating collocation points more selectively.
The residual-based adaptive resampling (RAR) method~\cite{DeepXDE2021,Rigas2024} allocates more sampling points to regions with larger residuals. 
Subsequent studies further incorporated residual distribution information to balance sampling between regions with large and small residuals~\cite{Mao2023,FIAS2023,Nabian2021}. Neural networks have also been used to learn adaptive sampling strategies~\cite{DAS2023}. Although these methods can improve solution accuracy or sampling efficiency, they often require additional residual evaluation, error estimation, or auxiliary training procedures, which may introduce extra computational overhead.

An alternative approach is to reduce the sampling burden by decomposing multi-scale problems and treating different scales separately.
Several multi-scale frameworks have been proposed~\cite{Aldirany2024,Wang2024,Song2025,LingHu2025}, where networks are trained to learn residual corrections, scale-dependent components, or higher-order terms across different scales.
While these methods have achieved notable progress in handling multi-scale nonlinear PDEs, they usually require additional network modules, scale-separation procedures, or coupled training stages.
In the computer vision field, where large datasets are common in training video models, Wu et al.~\cite{Facebook2020} introduced a multigrid method to accelerate training by alternating between spatial-temporal resolutions.
A similar strategy has also been applied to super-resolution video models~\cite{Lin2023}, where multi-resolution data are leveraged to improve the training efficiency and performance of a single model.

Motivated by these observations, we introduce a multi-resolution training strategy into the physics-informed machine learning framework to reduce the cost associated with high-resolution residual sampling for multi-scale PDE problems. Instead of constructing additional scale-decomposition models, the proposed strategy dynamically adjusts the residual sampling resolution during training, allowing the model to exploit the efficiency of low-resolution sampling while retaining the accuracy benefits of high-resolution refinement. Based on this idea, we propose a multi-resolution training-enhanced PIKAN framework, referred to as MR-PIKAN, for solving forward and inverse PDE problems with multi-scale features. The main contributions of this study are summarized as follows:
\begin{itemize}
    \item We analyze the training dynamics of PIKAN models under various single-resolution training settings from the perspective of information bottleneck theory. The results confirm that high-resolution sampling is essential for accurately resolving multi-scale physical systems.
    \item We introduce a simple yet effective multi-resolution training strategy within the PIKAN framework, resulting in MR-PIKAN, where training at lower resolutions effectively reduces the overall training cost.
    \item We test the proposed MR-PIKAN model on a series of multi-scale forward and inverse PDE problems. The results indicate that MR-PIKAN can effectively enhance computational efficiency without compromising accuracy.
\end{itemize}
The rest of this paper is organized as follows.
In Section~\ref{sec:PIML}, we present a brief overview of the physics-informed machine learning formulation. Subsequently, the multi-resolution training method is presented in Section~\ref{sec:MR}, and the MR-PIKAN framework is introduced in detail.
Next, several numerical examples are tested in Section~\ref{sec:Numericalexamples} to demonstrate the accuracy and efficiency of the proposed MR-PIKAN.
Finally, the main conclusions of this study are summarized in Section~\ref{sec:Conclusions}.

\section{Physics-informed machine learning}
\label{sec:PIML}
\subsection{Physics-informed machine learning framework}
In this section, we present a brief overview of the physics-informed machine learning (PIML) framework for solving PDE problems~\cite{raissiPhysicsinformedNeuralNetworks2019,Karniadakis2021}.
We consider the following partial differential equation:
\begin{equation}\label{eq:staticprob}
	\left\{
	\begin{aligned}
		\mathcal{D}\left[ \boldsymbol{u} \right] \left( \boldsymbol{x} \right) &= \boldsymbol{f}\left( \boldsymbol{x} \right)
		&&\textrm{in } \Omega \\
		\mathcal{B}\left[ \boldsymbol{u} \right] \left( \boldsymbol{x} \right) &= \boldsymbol{g}\left( \boldsymbol{x} \right)
		&&\textrm{on } \partial \Omega
	\end{aligned}
	\right. 
\end{equation}
where $\mathcal D \left[ \cdot \right]$ represents the differential operator acting in the problem domain $\Omega$, and $\boldsymbol{f} \left( \boldsymbol{x} \right)$ denotes the source term. $\mathcal B \left[ \cdot \right]$ represents the boundary operator, and $\boldsymbol{g} \left( \boldsymbol{x} \right)$ specifies the prescribed boundary condition on $\partial \Omega$. The basic idea of PIML is to approximate the solution of Eq.~(\ref{eq:staticprob}) using a representation model ${\boldsymbol u}_{\boldsymbol \theta} \left( {\boldsymbol x} \right)$, where $\boldsymbol \theta$ denotes its trainable parameters.
Then, the optimal parameters $\boldsymbol \theta'$ can be obtained by minimizing the physics-informed loss function:
\begin{equation}\label{eq:loss}
	\begin{aligned}
	\mathcal L\left({\boldsymbol{\theta}}\right)
	&=\omega_r \mathcal L_r\left({\boldsymbol{\theta}}\right) + \omega_b \mathcal L_{b}\left({\boldsymbol{\theta}}\right)\\
	\mathcal{L}_r \left({\boldsymbol{\theta}}\right) &=\frac{1}{N_r}\sum_{i=1}^{N_r}{\left| \mathcal{D}\left[ \boldsymbol{u}_{\boldsymbol{\theta }} \right] \left( \boldsymbol{x}_{r}^{i} \right) -\boldsymbol{f}\left( \boldsymbol{x}_{r}^{i} \right) \right|^2} \\
	\mathcal{L}_b \left({\boldsymbol{\theta}}\right) &=\frac{1}{N_b}\sum_{i=1}^{N_b}{\left| \mathcal{B}\left[ \boldsymbol{u}_{\boldsymbol{\theta }} \right] \left( \boldsymbol{x}_{b}^{i} \right) - \boldsymbol{g}\left( \boldsymbol{x}_{b}^{i} \right) \right|^2}
	\end{aligned}
\end{equation}
where $\mathcal{L}_r$ and $\mathcal{L}_b$ represent the loss terms associated with the governing equation and the boundary conditions, respectively. $N_r$ and $N_b$ denote the sizes of the training data $\boldsymbol{x}_{r}^{i}$ and $\boldsymbol{x}_{b}^{i}$, respectively.
The weight coefficients $\omega_r$ and $\omega_b$ are introduced to balance the contributions of different loss components.

\subsection{Physics-informed neural networks}
\begin{figure}[ht]
	\centering
	\includegraphics[width=\textwidth]{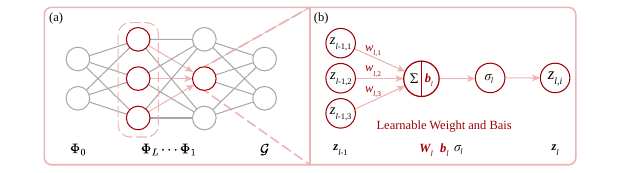}
	\caption{
		\textbf{Multi-layer perceptron (MLP).}
		(a) Architecture of an MLP.
		(b) Operation in the $l$-th hidden layer of an MLP, where $\boldsymbol{W}_l$ and $\boldsymbol{b}_l$ are learnable parameters and $\sigma_l$ is a prescribed nonlinear activation function.
	}
	\label{fig:MLP}
\end{figure}

In the conventional PINNs framework, the unknown solution is typically approximated by an MLP.
As shown in Fig.~\ref{fig:MLP}(a), an MLP consists of an input layer $\Phi_0$, several hidden layers $\Phi_1 \dots \Phi_{L}$, and an output layer $\cal G$, where each neuron is connected to every neuron in its adjacent layers. 
Considering an MLP with $L$ hidden layers, the nonlinear transformation can be expressed in the following recursive form:
\begin{equation}\label{eqFNN}
	{\boldsymbol u}_{\boldsymbol{\theta}}\left( {\boldsymbol x}\right)=
	{\text{MLP}} \left( {\boldsymbol x} \right)
	={\cal G} \circ {\Phi _L} \circ  \cdots  \circ {\Phi _1}\circ {\Phi _0}\left(  {\boldsymbol x} \right)
\end{equation}

For the $l$-th hidden layer, there are $N_l$ neurons with the activation function $\sigma_l$.
Thus, as illustrated in Fig.~\ref{fig:MLP}(b), the transformation ${\Phi _l}\left( \cdot \right)$ takes the form:
\begin{equation}\label{eqhiddenlayer}
	\begin{array}{*{20}{c}}
		\boldsymbol{z}_l = {\Phi _l}\left(  {\boldsymbol z}_{l-1} \right)
		=\sigma_l \left( {{\boldsymbol W}_l}{\boldsymbol z}_{l-1} + {\boldsymbol b}_l \right) &\text{for} &{l \in \left[1,L\right]}
	\end{array}
\end{equation}
where $\boldsymbol{z}_l = \left\{z_{l,1},\dots, z_{l,N_l}\right\}$ are the neural outputs of layer $l$.
The learnable weight matrix and bias vector of the $l$-th hidden layer are denoted by ${\boldsymbol W}_l\in {\mathbb R}^{{N_l} \times {N_{l-1}}}$ and ${\boldsymbol b}_l \in {\mathbb R}^{N_l}$, respectively.
For the input layer ${\Phi _0}\left( \cdot \right)$, we have 
\begin{equation}
	{\Phi _0}\left({\boldsymbol x} \right)
	={\boldsymbol x} \in {\mathbb R}^{d_{\rm in}}
\end{equation}
where the number of neurons in the input layer is set as $N_0=d_{\mathrm{in}}$ to match the input dimension of the problem.
For the output layer ${\cal G}\left(\cdot\right)$, linear mapping is typically adopted to match the dimension of the problem output.
Thus, we have
\begin{equation}
	{\cal G}\left({\boldsymbol z}_L \right)
	={{\boldsymbol W}_{L+1}}{\boldsymbol z}_L + {\boldsymbol b}_{L+1}
\end{equation}
The trainable parameters in the neural network 
$\text{MLP} \left( {\boldsymbol x} \right)$
are $\boldsymbol{\theta}=\left\{ {\boldsymbol W}_l,{\boldsymbol b}_l \mid l=1,\dots, L+1\right\}$.
Through the nonlinear activation function $\sigma_l$, the MLP establishes a highly nonlinear mapping between the input variable $\boldsymbol{x}$ and the network output $\boldsymbol{u}_{\boldsymbol{\theta}}(\boldsymbol{x})$.

\subsection{Physics-informed Kolmogorov--Arnold networks}
While MLPs have been widely adopted as universal function approximators, they typically construct nonlinear mappings through compositions of linear transformations and predefined activation functions.
An alternative perspective is provided by the Kolmogorov--Arnold representation theorem~\cite{Kolmogorov1957}, which demonstrates that multivariate continuous functions can be represented through compositions of univariate functions and summations. Specifically, for a continuous function $f:[0,1]^d\rightarrow \mathbb{R}$, the theorem states that
\begin{equation}\label{eq}
	f(x_1,\ldots,x_d)=\sum_{q=0}^{2d} \Phi_q\left(\sum_{p=1}^{d}\phi_{q,p}(x_p)\right)
\end{equation}
where $\phi_{q,p}$ and $\Phi_q$ are univariate continuous functions. This theorem indicates that, in principle, a multivariate mapping can be constructed from learnable one-dimensional transformations combined through additive operations.
	
Motivated by this theoretical insight, recent studies have revisited KANs as promising alternatives to MLPs, where learnable univariate functions are incorporated into neural network architectures~\cite{KAN2024a}. The recent work of Liu et al.~\cite{KAN2024a} established a generalized KAN framework with arbitrary widths and depths, providing a flexible architecture that has motivated subsequent developments. Based on this architecture, Physics-Informed Kolmogorov--Arnold Networks (PIKANs)~\cite{FAIRKAN2024,wangKolmogorovArnoldInformed2025} have been introduced, in which the solution representation is defined by a KAN.

\begin{figure}[ht]
	\centering
	\includegraphics[width=\textwidth]{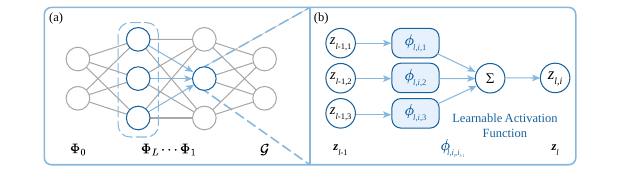}
	\caption{
		\textbf{Kolmogorov--Arnold network (KAN).}
		(a) Architecture of a KAN.
		(b) Operation of a KAN layer, where the univariate nonlinear activation functions $\phi_{l,i_l,i_{l-1}}$ are learnable.
	}
	\label{fig:KAN}
\end{figure}

As illustrated in Fig.~\ref{fig:KAN}(a), the nonlinear transformation of a KAN with $L$ KAN layers can be expressed in the following recursive form:
\begin{equation}\label{eqKAN}
	{\boldsymbol u}_{\boldsymbol{\theta}}\left( {\boldsymbol x}\right)=
	{\text{KAN}} \left( {\boldsymbol x}\right)
	={\cal G} \circ {\Phi_{L}} \circ \cdots  \circ {\Phi _1} \circ {\Phi _0}\left( {\boldsymbol x} \right)
\end{equation}
where ${\Phi _{l}}$ represents the nonlinear transformation of the $l$-th KAN layer.
The input layer ${\Phi _0}\left( \cdot \right)$ and output layer $\mathcal{G}\left( \cdot \right)$ in KAN are typically simplified to an identity mapping.

In the $l$-th KAN layer, there are $N_{l} \times N_{l-1}$ univariate activation functions $\phi_{l,i_{l},i_{l-1}}\left( \cdot \right)$.
Therefore, the transformation ${\Phi_{l}}\left( {\boldsymbol x} \right)$ of a KAN layer, as shown in Fig.~\ref{fig:KAN}(b), can be written as a function matrix:
\begin{equation}\label{eq:KANlayer}
	\boldsymbol{z}_{l}=\Phi _{l}\left( \boldsymbol{z}_{l-1} \right) =\left[ \begin{matrix}
		\phi _{l,1,1}\left( \cdot \right)	&\phi _{l,1,2}\left( \cdot \right)&\cdots&\phi _{l,1,N_{l-1}}\left( \cdot \right)\\
		\phi _{l,2,1}\left( \cdot \right)	&\phi _{l,2,2}\left( \cdot \right)&\cdots&\phi _{l,2,N_{l-1}}\left( \cdot \right)\\
		\vdots&\vdots&\ddots&\vdots\\
		\phi _{l,N_{l},1}\left( \cdot \right)&\phi _{l,N_{l},2}\left( \cdot \right)&\cdots&\phi _{l,N_{l},N_{l-1}}\left( \cdot \right)\\
	\end{matrix}
	\right]
	\left[ \begin{matrix}
		z_{l-1,1}\\
		z_{l-1,2}\\
		\vdots\\
		z_{l-1,N_{l-1}}\\
	\end{matrix}
	\right]
\end{equation}
where $\boldsymbol{z}_l = \left\{z_{l,1},\dots, z_{l,N_l}\right\}$ are the neural outputs of the $l$-th KAN layer.
Here, $\phi_{l,i_{l},i_{l-1}}$ represents the learnable univariate activation function corresponding to the $i_{l}$-th feature of $\boldsymbol{z}_l$ and the $i_{l-1}$-th feature in $\boldsymbol{z}_{l-1}$.

Different KAN variants are mainly determined by the specific functional form of $\phi(x)$.
In the generalized KAN formulation of Liu et al.~\cite{KAN2024a},
$\phi\left(x\right)$ is defined as a weighted combination of a basis function $b\left(x\right)$ and B-splines.
In particular:
\begin{equation}\label{eq:vanillaKANlayer}
	\begin{aligned}
		\phi \left( x \right) 	&=\omega _bb\left( x \right) +\omega _s\text{spline}\left( x \right) \\
		b\left( x \right) 		&=\text{silu}\left( x \right) =\frac{x}{1+e^{-x}}\\
		\text{spline}\left( x \right) &=\sum_i{c_iB_i\left( x \right)}
	\end{aligned}
\end{equation}
where $\omega_{b}, \omega_{s}$ and $c_i$ are trainable parameters.
This formulation has motivated a series of KAN variants in which the learnable univariate functions are represented using different basis functions, such as radial basis functions~\cite{RBFKAN}, wavelets~\cite{WaveletKAN}, and Jacobi polynomials~\cite{ChebyKAN}.

In this study, we adopt Chebyshev KAN, a KAN variant in which the learnable univariate activation functions are parameterized by Chebyshev polynomial expansions. Compared with the vanilla KAN model, Chebyshev KAN has been reported to achieve higher computational efficiency~\cite{ChebyKAN}. 
Specifically, each $\phi\left(x\right)$ is represented as a weighted linear combination of Chebyshev polynomials up to degree $n$, and the input $x$ is normalized to the range $\left(-1, 1\right)$ using a hyperbolic tangent function:
\begin{equation}\label{eq:ChebyKANlayer}
	\begin{aligned}
		\phi \left( x \right) &=\sum_{k=0}^{n}{\Theta_k T_k\left( \tilde{x} \right)}\\
		\tilde{x} &= \tanh \left(x\right)
	\end{aligned}	
\end{equation}
where $\Theta_k$ denotes the learnable coefficient associated with the $k$-th Chebyshev polynomial. The Chebyshev polynomials of the first kind can be evaluated using the following explicit expression:
\begin{equation}\label{eq:Tnk}
	T_n\left( x \right) = \cos \left(n \arccos\left( x \right)\right)
\end{equation}
The explicit analytical form of Chebyshev polynomials enables direct and efficient evaluation of the activation functions during forward propagation, which becomes increasingly advantageous for higher-order polynomial expansions.
As reported in~\cite{wangKolmogorovArnoldInformed2025}, PIKANs exhibit reduced spectral bias compared with MLP-based PINNs and offer certain advantages in modelling multi-scale problems.
Therefore, in this paper, we focus on using PIKAN models based on the Chebyshev KAN representation to solve multi-scale PDE problems.

\section{Multi-resolution training method}
\label{sec:MR}
Once an appropriate neural-network representation is selected, the PIML model is trained by optimizing its parameters so that the represented solution satisfies the governing equation and the prescribed constraints.
Training a PIML model involves minimizing the loss function $\mathcal{L}$ which quantifies the disagreement between the approximation of the representation model and the solution of the physical system.
In practice, the loss function $\mathcal{L}$ is approximated over $N_r$ residual points sampled from the domain $\Omega$ and $N_b$ boundary points sampled from $\partial \Omega$.

\begin{figure}[h]
    \centering
 	\includegraphics[width=0.8\textwidth]{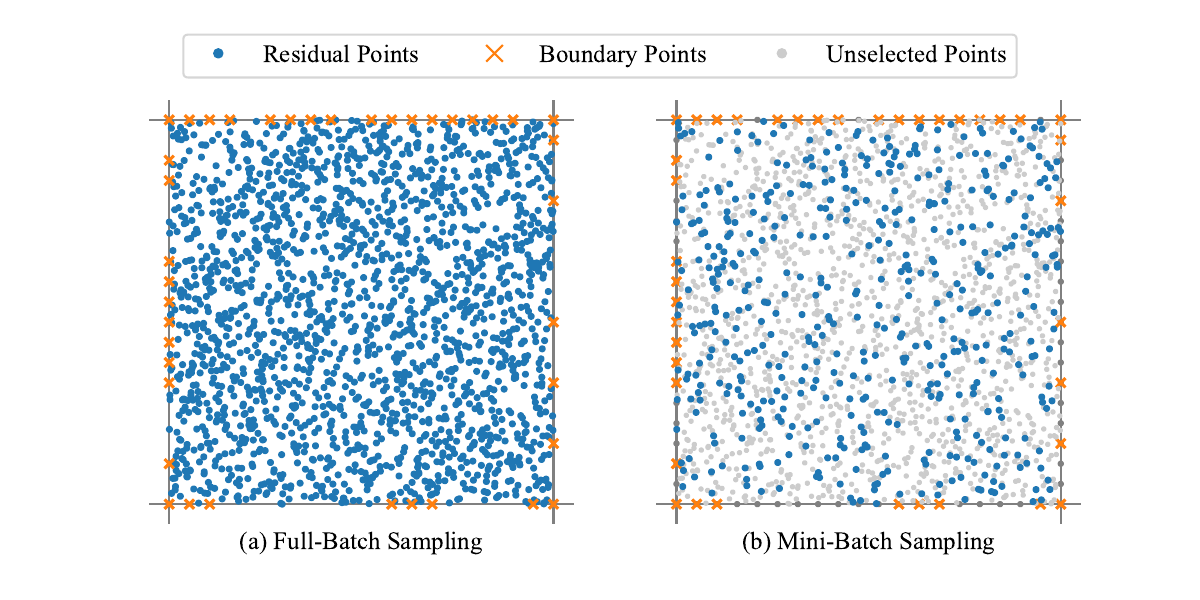}
    \caption{\textbf{Two single-resolution (SR) sampling strategies for residual points in a computational domain.}
    	(a) Full-batch sampling: the same large set of residual points (blue dots) is used at every iteration, which improves the stability of gradient evaluation but increases the computational cost.
    	(b) Mini-batch sampling: a randomly selected subset of residual points (blue dots) is used at each iteration, while the remaining points (gray dots) are temporarily excluded from gradient evaluation. This strategy reduces the computational burden but may introduce stochastic noise into the optimization process.
    	In both strategies, a relatively small set of boundary points (orange crosses) is typically used in full when computing the boundary-related loss terms.
    }
    \label{fig:batch_DP}
\end{figure}

As shown in Fig.~\ref{fig:batch_DP}, the residual points can either be sampled once throughout the entire training process or sampled individually at each iteration, leading to full-batch sampling and mini-batch sampling, respectively.
Regardless of the batch sampling strategy used, sufficiently dense sampling is necessary to effectively capture the multi-scale features of the system~\cite{Wang2024}.
To the best of the authors' knowledge, the batch size of PIML models in previous studies typically remains fixed throughout the entire training process.
Therefore, we refer to this sampling method with a fixed batch size as the single-resolution (SR) training strategy.

\subsection{Single-resolution training dynamics}
\label{subsec:SR}
The batch size of the SR training strategy has a significant impact on the accuracy of the results, with this effect becoming increasingly pronounced when solving multi-scale physical problems.
Recently, the information bottleneck (IB) principle has been employed to analyze the training dynamics of PIML models.
We first analyze the accuracy and training dynamics of a forward problem with multi-scale features under different single-resolution settings.

Consider the Poisson equation in 2D as follows:
\begin{equation}\label{eq:2Dpoisson}
	\left\{
	\begin{aligned}
		-\Delta u\left( x,y \right) &= f\left( x,y \right) 	&&\textrm{in } \Omega\\
		u\left( x,y \right) 		&= g\left( x,y \right)	&&\textrm{on } \partial\Omega
	\end{aligned}
	\right.
\end{equation}
where $\left(x,y\right)\in \Omega =\left[-1,1\right]\otimes \left[-1,1\right]$, and $g(x,y)=0$ is prescribed on $\partial\Omega$.
In this example, we select the exact solution with multi-scale features as:
\begin{equation}\label{eq:2DpoissonExact}
	u_{\rm exact}\left(x,y\right)= \sin\left(2\pi x\right) \sin\left(4\pi y\right)+\frac{1}{2} \sin\left(6\pi x\right) \sin\left(8\pi y\right)
\end{equation}
The network is trained by minimizing the loss function given below:
\begin{equation}\label{eq:2DPoissonLoss}
	\begin{aligned}
		\mathcal L\left({\boldsymbol{\theta}}\right)&=\omega_r \mathcal L_r\left({\boldsymbol{\theta}}\right) + \omega_b \mathcal L_{b}\left({\boldsymbol{\theta}}\right)\\
		&=\frac{\omega_r}{N_r}\sum_{i=1}^{N_r}{\left| \Delta u_{\boldsymbol{\theta }}\left( x_r^i,y_r^i \right) +f\left( x_r^i,y_r^i \right) \right|^2} +\frac{\omega_b}{N_b}\sum_{i=1}^{N_b}{\left| u_{\boldsymbol{\theta }}\left(x_b^i,y_b^i\right)-g\left( x_b^i,y_b^i \right) \right|^2}\\
	\end{aligned}
\end{equation}

We then train PIML models with different neural-network approximators using a single-resolution training strategy. The MLP-based and KAN-based models are denoted as SR-PINN and SR-PIKAN, respectively. For a fair comparison, SR-PINN and SR-PIKAN are configured with comparable numbers of trainable parameters for this problem.
Further details on the network architecture and hyperparameter settings can be found in~\ref{sec:appendixA}.
In this forward problem example, we select $N_b=800$ sampling points on the boundary and vary the number of residual points within the computational domain as $N_r=$320, 640, 1280, 2560 and 5120.

As shown in Table~\ref{tab:L2fullbatch}, both SR-PINN and SR-PIKAN achieve improved accuracy as the sampling resolution increases, confirming the importance of sufficient residual points for resolving the multi-scale solution in Eq.~\ref{eq:2DpoissonExact}. Under the same resolution setting, SR-PIKAN generally yields smaller $H_1$ errors than SR-PINN, indicating its advantage in capturing not only the solution field but also its first-order derivatives.
\renewcommand{\arraystretch}{1.25}
\begin{table}[H]
	\footnotesize
	\centering
	\begin{tabular}{llccccc}
		\toprule
		\multicolumn{2}{c}{Single-resolution setting} & $N_r=320$ & $N_r=640$ & $N_r=1280$ & $N_r=2560$ & $N_r=5120$ \\
		\midrule
		\multirow{2}{*}{$L_2{\text{-error}}$}& SR-PIKAN &9.195e-01 &7.545e-01 & 4.685e-01 & 2.048e-01 & 1.945e-02 \\
		& SR-PINN &4.068e+00 &2.143e+00 & 1.990e+00 & 1.030e-01 & 3.133e-02 \\
		\midrule
		\multirow{2}{*}{$H_1{\text{-error}}$} & SR-PIKAN
		& 8.189e-01
		& 6.698e-01
		& 3.880e-01 
		& 8.693e-02 
		& 4.819e-03 \\
		
		& SR-PINN 
		& 1.663e+00 
		& 1.232e+00 
		& 7.229e-01 
		& 5.154e-02 
		& 2.556e-02 \\
		\bottomrule
	\end{tabular}
	\caption{\textbf{Relative $L_2$ and $H_1$ errors of SR-PIKAN and SR-PINN models under different single-resolution settings.}
		Both SR-PIKAN and SR-PINN models achieve more accurate results as the sampling resolution increases.
	}
	\label{tab:L2fullbatch}
\end{table}
\renewcommand{\arraystretch}{1}
\begin{figure}[h]
	\centering
	\includegraphics[width=\textwidth]{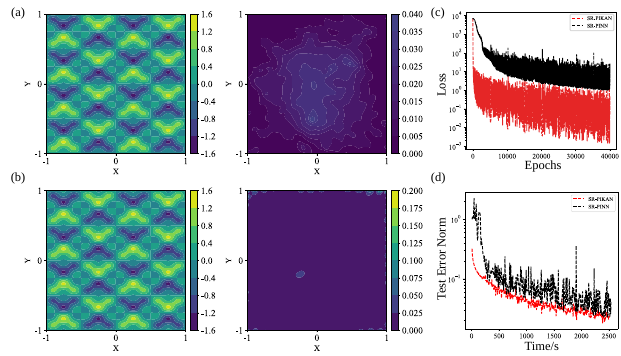}
	\caption{\textbf{Results of SR-PIKAN and SR-PINN models trained with the single-resolution setting of $\boldsymbol{N_r=5120}$}. 
		(a) Predicted solution and point-wise error of the SR-PIKAN model.
		(b) Predicted solution and point-wise error of the SR-PINN model.
		(c) Training loss $\mathcal L$ over epochs for the SR-PIKAN and SR-PINN models.
		(d) Test $L_2{\text{-error}}$ over training time for the SR-PIKAN and SR-PINN models.
	}
	\label{fig:SR-poisson}
\end{figure} 
The SR-PIKAN and SR-PINN results obtained under the single-resolution setting of $N_r=5120$ are shown in Fig.~\ref{fig:SR-poisson}(a) and (b), respectively.
The results show that, with sufficient sampling, both the SR-PINN and SR-PIKAN models can accurately approximate the solution of the problem with multi-scale features.
Fig.~\ref{fig:SR-poisson}(c) and (d) illustrate the evolution of the training loss and $L_2{\text{-error}}$ throughout the training process.
These results indicate that, after sufficient training, both SR-PINN and SR-PIKAN exhibit comparable test accuracy.

To further understand the training behavior, we examine the empirical neural tangent kernel (NTK) spectra of the SR-PINN and SR-PIKAN models. For the loss formulation considered in Eq.~\ref{eq:2DPoissonLoss}, the corresponding weighted empirical NTK can be written in the following block form:
\begin{equation}\label{eq:ntk_block_form}
	\mathbf{K}_{\mathrm{NTK}}
	=
	\begin{bmatrix}
		\frac{w_b}{N_b}\mathbf{J}_{b}\mathbf{J}_{b}^{T}
		&
		\frac{w_r}{N_r}\mathbf{J}_{b}\mathbf{J}_{r}^{T}
		\\
		\frac{w_b}{N_b}\mathbf{J}_{r}\mathbf{J}_{b}^{T}
		&
		\frac{w_r}{N_r}\mathbf{J}_{r}\mathbf{J}_{r}^{T}
	\end{bmatrix}
\end{equation}
where $\mathbf{J}_{b}$ and $\mathbf{J}_{r}$ denote the Jacobians of $u_{\boldsymbol{\theta}}\left(\boldsymbol{x}_b\right)$ and $\Delta u_{\boldsymbol{\theta}}\left(\boldsymbol{x}_r\right)$ with respect to the trainable parameters $\boldsymbol{\theta}$, respectively. Further details on the NTK analysis are provided in~\ref{sec:appendixNTK}.
\begin{figure}[h!]
	\centering
	\includegraphics[width=\textwidth]{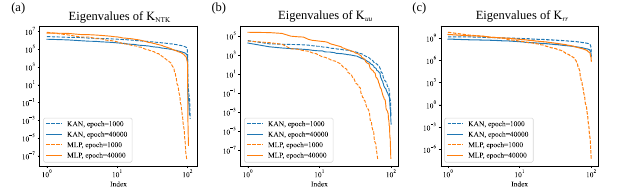}
	\caption{
		\textbf{NTK eigenvalue spectra of SR-PIKAN and SR-PINN models}. 
		(a) Eigenvalues of the weighted NTK dynamics matrix $\mathbf{K}_{\mathrm{NTK}}$.
		(b) Eigenvalues of the boundary NTK block $\mathbf{K}_{uu}$.
		(c) Eigenvalues of the residual NTK block $\mathbf{K}_{rr}$.
	}
	\label{fig:NTK_SR}
\end{figure} 

As shown in Fig.~\ref{fig:NTK_SR}, although the SR-PINN and SR-PIKAN models can achieve comparable final errors after sufficient training, their NTK spectra exhibit clear differences. Compared with SR-PINN, SR-PIKAN shows a slower eigenvalue decay and retains larger effective eigenvalues in the high-index region, especially for the residual NTK block $\mathbf{K}_{rr}$. This indicates that the SR-PIKAN architecture provides more balanced convergence modes and alleviates the rapid spectral decay observed in SR-PINNs. From the NTK perspective, these results indicate that SR-PIKAN provides improved trainability for PDE residual minimization and more effective error reduction across multi-scale solution components.

Furthermore, we analyzed the training dynamics of SR-PIKAN models at different single resolutions from the perspective of IB theory.
In the framework of IB theory, a commonly used metric to quantify training dynamics of a PIML model is the batch-wise signal-to-noise ratio (SNR), which is defined as:
\begin{equation}\label{eq:SNR}
	\mathrm{SNR} = \frac{\|\mu\|_2}{\|\sigma\|_2} = \frac{\|\mathbb{E}[\nabla_\theta \mathcal{L}_\mathcal{B}]\|_2}{\| \mathrm{std}[\nabla_\theta \mathcal{L}_\mathcal{B}] \|_2}
\end{equation}
where $\|\mu\|_2$ and $\|\sigma\|_2$ denote the $L_2$ norms of the batch-wise mean and standard deviation of $\nabla_\theta \mathcal{L}_\mathcal{B}$ for all network parameters.
The study in~\cite{anagnostopoulosLearningPINNsPhase2026} suggests that it is during the total diffusion phase when the neural network achieves optimal convergence.
Furthermore, the gradient homogeneity metric, $\textrm{SRR}_\mathcal{B}$, can be used to identify the point at which the model enters the total diffusion phase.
\begin{equation}\label{eq:SRR_B}
	\mathrm{SRR}_\mathcal{B} = \|\mathbb{E}[\nabla_\theta \mathcal{L}_\mathcal{B}]\|_2 / \|\sqrt{\mathbb{E}[(\nabla_\theta \mathcal{L}_\mathcal{B})^2]} \|_2
\end{equation}
The SNR is tightly linked to the gradient homogeneity metric $\textrm{SRR}_\mathcal{B}$ among samples through the following expression:
\begin{equation}\label{eq:SNR_SRR}
	\mathrm{SNR} = \frac{|\mathrm{SRR}_\mathcal{B}|}{\sqrt{1 - \mathrm{SRR}_\mathcal{B}^2}}
\end{equation}
Further details on the training dynamics can be found in reference~\cite{anagnostopoulosLearningPINNsPhase2026}.

\begin{figure}[H]
	\centering
	\includegraphics[width=\textwidth]{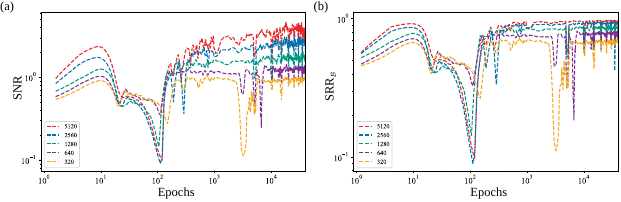}
	\caption{\textbf{Training dynamics of SR-PIKAN models trained under different single-resolution settings.} 
		(a) Signal-to-noise ratio (SNR) of SR-PIKAN models trained under different single-resolution settings.
		(b) Gradient homogeneity metric SRR$_\mathcal{B}$ of SR-PIKAN models trained under different single-resolution settings. All models entered the total diffusion phase after 1000 epochs, during which models trained with higher-resolution sampling exhibited a higher signal-to-noise ratio.
	}
	\label{fig:SR}
\end{figure}

As shown in Fig.~\ref{fig:SR}, all five SR-PIKAN models entered the total diffusion phase after 1000 epochs, during which the gradient homogeneity metric $\textrm{SRR}_\mathcal{B}$ remained at a relatively high level close to 1.0.
At this stage, the SNR of each SR-PIKAN model also remains stable, with higher-resolution models demonstrating higher SNR values.
The results indicate that for physical fields such as Eq.~\ref{eq:2DpoissonExact}, which exhibit multi-scale features, high-resolution sampling can effectively improve the signal-to-noise ratio of the optimization gradients. This helps stabilize the training process and provides a favorable condition for achieving accurate solutions.

\subsection{Multi-resolution training strategies}
Although the single-resolution training results for the benchmark problem show that higher-resolution sampling improves both training and test accuracy, it also significantly increases the computational cost.
Inspired by multi-grid methods in traditional numerical analysis and the solution of PDEs~\cite{Hackbusch}, a multi-resolution (MR) strategy has been introduced for training PIML models.
We utilize sampling point sets with multiple resolutions to train the network sequentially or alternately, thereby reducing the training time of the model while ensuring accuracy.

Consider the following parameter update process based on gradient descent: 
\begin{equation}\label{eq:update_operator}
	\boldsymbol{\theta}^{t+1} = \boldsymbol{\theta}^t - \eta \cdot \mathcal{G}^t_n
\end{equation}
where $\mathcal{G}^t_n$ denotes the stochastic approximation of the gradient of the loss function $\mathcal{L}_n(\boldsymbol{\theta}^t)$ at epoch $t$, computed using $n$ sample points.
The learning rate $\eta$ determines the update step size and is selected empirically or through hyperparameter tuning to promote stable convergence during training.
For the single-resolution training strategy described in Section~\ref{subsec:SR}, the batch size $n=N_{\text{batch}}$ is fixed during the training process.
Thus, we have:
\begin{equation}\label{eq:Gn_SR}
		\mathcal{G}^t_n=\mathcal{G}^t_{N_\text{batch}}=\frac{1}{N_{\text{batch}}} \sum_{i=1}^{N_{\text{batch}}}{ \nabla_{\boldsymbol{\theta}} \mathcal{L}_{i}(\boldsymbol{\theta}^t)}		
\end{equation}
The algorithm for the multi-resolution strategy is shown in Algorithm~\ref{alg:MR}, where the stochastic gradient estimate $\mathcal{G}^t_n$ depends on the training resolution function $N\left( t \right)$ at iteration step $t$.
At this point, the stochastic approximation of the gradient takes the form:
\begin{equation}\label{eq:Gn_MR}
	\mathcal{G}^t_n=\mathcal{G}^t_{N\left( t \right)}=\frac{1}{N\left( t \right)} \sum_{i=1}^{N\left( t \right)}{ \nabla_{\boldsymbol{\theta}} \mathcal{L}_{i}(\boldsymbol{\theta}^t)}	
\end{equation}

\begin{algorithm}
	\caption{Multi-resolution training strategy}
	\label{alg:MR}
	\begin{algorithmic}[1]
		\Require Training iterations $T$, training resolution function $N \left(t\right)$
		\State \textbf{Initialization:} Xavier normal scheme: $\boldsymbol{\theta}^0$
		\For{Training iteration $t=1,\cdots, T$}
		\State \textbf{Approximate the gradient}: $\mathcal{G}^t_n$ using Eq.~\ref{eq:Gn_MR}
		\State \textbf{Update parameters}: $\boldsymbol{\theta}^{t+1} = \boldsymbol{\theta}^t - \eta \cdot \mathcal{G}^t_n$
		\EndFor
	\end{algorithmic}
\end{algorithm}

\begin{figure}[h]
	\centering
	\includegraphics[width=\textwidth]{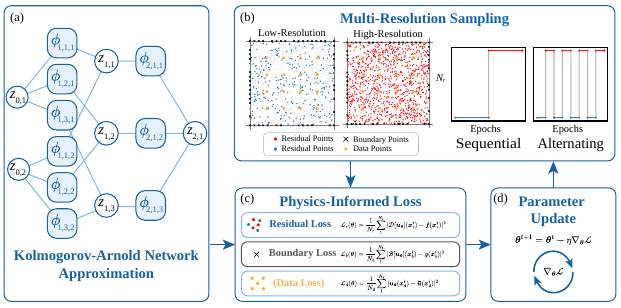}
	\caption{\textbf{Multi-resolution training-enhanced Kolmogorov--Arnold networks.}
	(a) KAN-based neural network approximation for the target variable.
	(b) Multi-resolution sampling is conducted based on a predefined strategy, where the number of residual points is varied sequentially or alternately across training stages.
	(c) Physics-informed loss combining PDE residual and boundary terms, with an optional data term when reference data are available.
	(d) Gradient-based parameter update followed by the next training epoch.
	}
	\label{fig:MR-PIKAN}
\end{figure}

Based on the above multi-resolution training strategy, we propose a multi-resolution training-enhanced PIKAN framework, referred to as MR-PIKAN. The overall workflow of the proposed MR-PIKAN framework is illustrated in Fig.~\ref{fig:MR-PIKAN}. The target solution is first approximated by a KAN-based neural representation. During training, the physics-informed loss is constructed from the PDE residual, boundary conditions, and, when available, additional data constraints. Different from conventional single-resolution training, the number of residual points is dynamically adjusted according to a predefined multi-resolution sampling strategy, either sequentially or alternately across training stages. Specifically, in the sequential multi-resolution strategy, the model is first trained at a lower resolution for a prescribed number of epochs and then refined at a higher resolution. In the alternating multi-resolution strategy, the model is trained alternately at different resolutions until the end of training. The resulting loss gradients are then used to update the network parameters, and the updated model proceeds to the next training stage. In this way, MR-PIKAN combines the representation capability of KANs with multi-resolution sampling to improve the training efficiency for multi-scale PDE problems.

In the numerical implementation, the multi-resolution strategy is realized by varying the number of residual sampling points used to evaluate the PDE residual loss during training.
For forward problems, the network is trained using full-batch sampling points at different resolutions. 
For inverse problems, however, since a relatively large dataset is required for effective inference, mini-batch sampling is adopted, and the batch size is varied according to the prescribed resolution schedule. When the low-resolution stage uses a sufficiently small number of residual points, the computational cost of both the forward and backward passes can be substantially reduced, thereby improving the overall training efficiency.
In Section~\ref{sec:Numericalexamples}, we apply the proposed MR-PIKAN model to a series of forward and inverse problems with multi-scale features to demonstrate its effectiveness and efficiency.

\section{Numerical examples}
\label{sec:Numericalexamples}
In this section, we demonstrate the performance of the proposed MR-PIKAN model in solving multi-scale forward and inverse problems.
For implementation, we developed the PIKAN and PINN frameworks through the PyTorch~\cite{pytorch2019} library in Python.
Throughout all benchmarks, we chose the hyperbolic tangent as the activation function for the MLP and as the normalizing function for the Chebyshev KAN.
The network was initialized using the Xavier normal scheme~\cite{Xavier2010}.
All networks were trained using the Adam optimizer~\cite{kingma2015adam} with the default setting on a single NVIDIA RTX 4070 Ti GPU.
In particular, we employed exponential learning rate decay with a decay rate of 0.9 every 5,000 training iterations. More details are provided in~\ref{sec:appendixA}.

The primary metric used to evaluate the model's generalization ability is the relative $L_2$ error on the test dataset, which is expressed as:
\begin{equation}\label{eq:testL2}
	L_2{\text{-error}} = \frac{\Vert \boldsymbol{u}_{\boldsymbol{\theta }}\left( \boldsymbol{x} \right) -\boldsymbol{u}_{\text{exact}}\left( \boldsymbol{x} \right) \Vert_{2}}{\Vert \boldsymbol{u}_{\text{exact}}\left( \boldsymbol{x} \right) \Vert_{2}}
\end{equation}
where ${\boldsymbol u}_{\boldsymbol \theta} \left( {\boldsymbol x} \right)$ and ${\boldsymbol u}_{\text{exact}} \left( {\boldsymbol x} \right)$ represent the model approximation and the exact solution, respectively.
For forward problems with available exact solutions, we further report the relative $H_1$ error to assess the accuracy of both the solution field and its first-order derivatives. The relative $H_1$ error is defined as
	\begin{equation}\label{eq:testH1}
		H_1{\text{-error}} =
		\frac{
			\left(
			\Vert \boldsymbol{u}_{\boldsymbol{\theta}}-\boldsymbol{u}_{\text{exact}}\Vert_2^2
			+
			\Vert \nabla \boldsymbol{u}_{\boldsymbol{\theta}}-\nabla \boldsymbol{u}_{\text{exact}}\Vert_2^2
			\right)^{1/2}
		}{
			\left(
			\Vert \boldsymbol{u}_{\text{exact}}\Vert_2^2
			+
			\Vert \nabla \boldsymbol{u}_{\text{exact}}\Vert_2^2
			\right)^{1/2}
		}.
		\end{equation}
where $\nabla$ denotes the gradient operator.
A relatively dense test grid is used to obtain a more reliable estimate of the reported errors.

\subsection{2D Poisson equation}
\label{subsec:2D Poisson}
\subsubsection{Multi-resolution training results}
\begin{figure}[H]
	\centering
	\includegraphics[width=\textwidth]{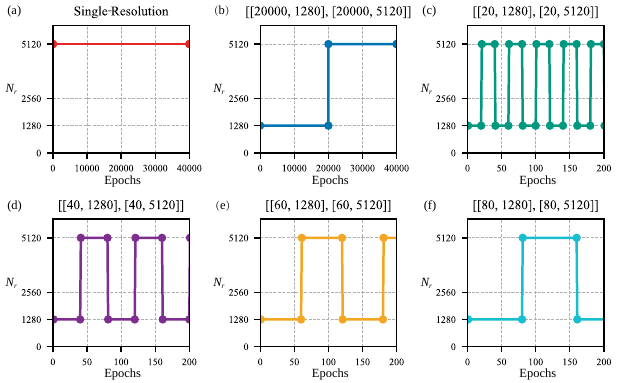}
	\caption{\textbf{Different training strategies.}
		(a) Single-resolution training strategy: $N_r$ remains constant throughout the entire training process.
		(b)-(f) Multi-resolution training strategies: The notation $[n,N_r]$ denotes a configuration where $N_r$ points are used for $n$ epochs before switching to the next configuration.
	}
	\label{fig:MRS}
\end{figure}

First, we apply MR-PINNs and MR-PIKANs to the problem described in Section~\ref{subsec:SR}, where the governing equation, boundary conditions, and analytical solution remain the same as previously specified.
We adopt the same PINN and PIKAN network configurations as those used in Section~\ref{subsec:SR}, so that the two models have approximately the same number of trainable parameters. 
Specifically, the MLP used in the PINN model consists of 6 hidden layers with 72 neurons per layer and uses the hyperbolic tangent activation function, resulting in 27,433 trainable parameters.
The KAN used in the PIKAN model has 2 hidden layers, each with 60 neurons, and utilizes Chebyshev polynomials of up to degree 6 for activation, with a total of 26,700 trainable parameters.
The boundary conditions are imposed using the same set of $N_b=800$ boundary sampling points as in Section~\ref{subsec:SR}.
In addition, we set the weight coefficients to $\omega_r=0.1, \omega_{b}=100$ for a more stable training.
A dense test set is used to evaluate the model accuracy in resolving multi-scale solution features.
The initial learning rate of PINNs and PIKANs is set to 1e-3 for this benchmark problem.
All results after 40,000 iterations are used for comparison across all examples.
As shown in Fig.~\ref{fig:MRS}(a), the single-resolution strategy keeps the number of residual points fixed at $N_r=5120$ throughout training. The corresponding results are referred to as SR-PINN or SR-PIKAN and are used for further comparison and discussion.

We first evaluate the sequential multi-resolution strategy by training the model successively at different resolutions, with all other training settings kept consistent with the single-resolution cases.
As shown by the blue curve in Fig.~\ref{fig:MRS}(b), the scheme [[20000,1280],[20000,5120]] represents 
sequential training for 20,000 steps at a low resolution $\left(N_r=1280\right)$, followed by another 20,000 steps at a high resolution $\left(N_r=5120\right)$, resulting in a total of 40,000 iterations.

\begin{figure}[H]
    \centering
    \includegraphics[width=\textwidth]{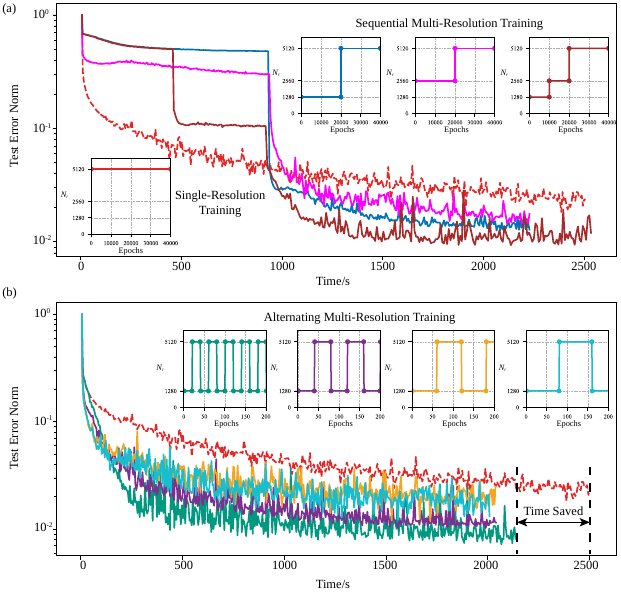}
    \caption{\textbf{Test-error evolution of MR-PIKAN models.}
    (a) Test-error evolution of sequential multi-resolution PIKAN models: The test errors of the MR-PIKAN models remain consistently lower than those of the single high-resolution model, represented by the red dashed line. During low-resolution training, a convergence plateau is clearly observed, where the test error cannot further decrease until it switches to a higher resolution.
    (b) Test-error evolution of alternating multi-resolution PIKAN models: The MR-PIKAN model achieves lower test errors compared with the SR-PIKAN model, while significantly reducing training time.
    }
    \label{fig:possionMR-KAN}
\end{figure}

 Fig.~\ref{fig:possionMR-KAN} shows the test $L_2$-error of the models trained using different multi-resolution training strategies as a function of time.
The red dashed lines in Fig.~\ref{fig:possionMR-KAN} represent the single-resolution training results obtained using a full-batch strategy at a resolution of 5,120, which serve as a reference for subsequent multi-resolution training.
The results show that models trained with multi-resolution strategies reduce training time while achieving comparable test accuracy, with occasional slight improvements.
The blue and magenta lines show that the models trained using the two-resolution strategy can reduce the training time by approximately 10\%, while the brown line represents the model trained using the three-resolution strategy, where the training time is nearly the same as that of the single-resolution case.
As shown in Fig.~\ref{fig:possionMR-KAN}(a), the model is first trained at low resolution, quickly reaching an accuracy plateau, with little further improvement observed before switching to a higher resolution.
Therefore, we further adopt an alternating multi-resolution training strategy, which switches between resolutions to reduce redundant training after the accuracy plateaus at a given resolution.
Since the model is trained using stochastic gradient descent at various resolutions, the iteration step can be treated as a hyperparameter.
 Fig.~\ref{fig:MRS}(c)-(f) illustrate the training strategies of the model when different alternating steps of 20, 40, 60 and 80 are selected, respectively.
 Fig.~\ref{fig:possionMR-KAN}(b) shows the test error of the models trained using the alternating multi-resolution training strategy with different step sizes.
The results show that all models trained using the alternating multi-resolution strategy save approximately one-sixth of the training time compared to the model trained using the single high-resolution strategy, while also achieving improved test accuracy compared to the single-resolution case.
The reduction in total training time is mainly attributed to the faster forward and backward passes at lower resolutions. 
Therefore, when the resolution increases to the point where the hardware can no longer support full-batch training, the alternating multi-resolution training strategy still offers potential for further improvement in training speed.
The results in Appendix~\ref{app:metrics} quantitatively compare the wall-clock training time of MR-PIKAN and SR-PIKAN, and further show that the multi-resolution training strategy does not introduce additional GPU memory consumption.
The above results indicate that using the alternating multi-resolution training strategy can effectively accelerate the training of the PIKAN model.
In addition, the alternating training strategy outperforms the sequential training strategy, and the results suggest that smaller iteration intervals lead to higher training efficiency.

\subsubsection{Training results of MR-PINNs and comparison across different models}
\begin{figure}[H]
	\centering
	\includegraphics[width=\textwidth]{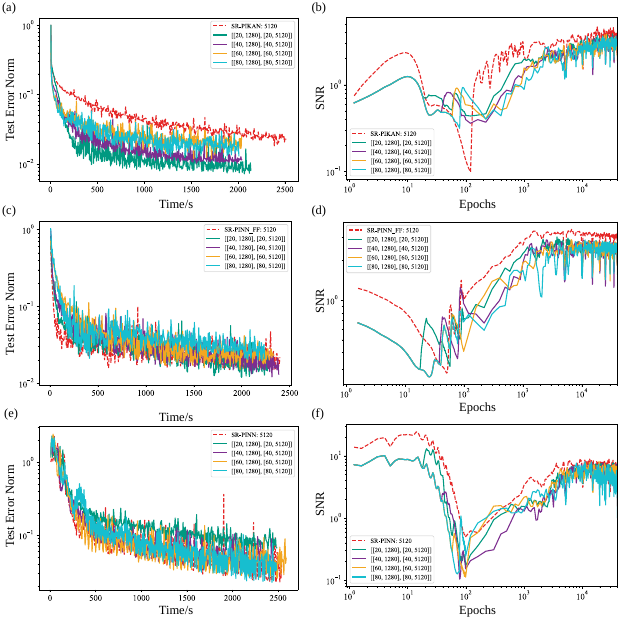}
	\caption{\textbf{Training dynamics of MR-PIKANs, MR-PINNs, and MR-PINNs with Fourier-feature embedding.}
		(a) Test-error evolution of MR-PIKAN models: A significant reduction in training time compared to the single-resolution model can be observed.
		(b) Training dynamics of MR-PIKAN models: Three typical phases can be observed: fitting, diffusion and total diffusion.
		(c) Test-error evolution of MR-PINN models with Fourier-feature embedding: The results obtained using single-resolution and multi-resolution training strategies for this type of model are similar in terms of both accuracy and training time. 
		(d) Training dynamics of MR-PINN models with Fourier-feature embedding: The models exhibit results similar to those for PIKAN models.
		(e) Test-error evolution of MR-PINN models: The vanilla PINN models, with a comparable number of trainable parameters, fail to achieve the same order of test accuracy as the previous two models on this problem.
		(f) Training dynamics of MR-PINN models: Vanilla PINN models exhibit a high SNR during the fitting phase; however, in the final total diffusion phase, the SNR fails to reach the values achieved during the initial phase.
	}
	\label{fig:MR-IB}
\end{figure}

Furthermore, we applied the same alternating multi-resolution training strategy to two types of PINN models two types of PINN models: vanilla PINNs and Fourier-feature PINNs.
The evolution of the test error and SNR during training for MR-PIKAN and the two PINN models is shown in Fig.~\ref{fig:MR-IB}.
The results in Fig.~\ref{fig:MR-IB}(a) and (b) show that MR-PIKAN models significantly reduce training time compared with the SR-PIKAN model.
Furthermore, the training of the PIKAN models exhibits three typical phases: fitting, diffusion, and total diffusion.
The results in Fig.~\ref{fig:MR-IB}(c) and (d) show that the Fourier-feature PINNs trained with the multi-resolution strategy achieve accuracy comparable to that of single-resolution training.
The training dynamics during the convergence process are similar to those of the PIKAN models.
However, no significant time savings are observed with the multi-resolution models.
As shown in Fig.~\ref{fig:MR-IB}(e) and (f), the vanilla PINN model, with a comparable number of trainable parameters, fails to achieve the same order of test accuracy as the previous two models on this problem.
These vanilla PINN models exhibit a high SNR during the fitting phase in the first 100 epochs.
However, they primarily capture lower-frequency components and become trapped in local minima.
In the total diffusion phase, the SNR of the PINN models remains lower than that observed in the initial fitting phase, mainly due to the spectral bias of the MLP architecture.

Overall, the analysis of the training dynamics of the three types of models shows that PIKAN and Fourier-feature PINN models achieve better accuracy and exhibit similar training behavior, whereas the vanilla PINN displays a distinctly different training pattern. This comparison further highlights the superior performance of MR-PIKAN in terms of both accuracy and efficiency for the considered forward problem with multi-scale features.
Moving forward, we employ the MR-PIKAN model to tackle challenging inverse problems with multi-scale features.

\subsection{Reaction-diffusion dynamics in a two-dimensional Gray--Scott model}
This example aims to highlight the performance of the proposed MR-PIKAN model in handling inverse problems, which are typically challenging due to their inherent ill-posedness and strong sensitivity to data quality~\cite{Zhang2025,Wu2025}.
Consider a two-dimensional Gray--Scott model~\cite{Gray1990} that describes a chemical reaction between two substances $U$ and $V$, both of which diffuse over time.
The Gray--Scott system is defined by two coupled equations that describe the behavior of two reacting substances:
\begin{equation}\label{eq:GrayScott}
	\begin{aligned}
		u_t &= \varepsilon_1 \Delta u +b\left(1-u\right)-uv^2\\
		v_t &= \varepsilon_2 \Delta v -dv + uv^2
	\end{aligned}
\end{equation}
where $\Delta$ is the Laplacian, and $u, v$ represent the concentrations of two reacting substances $U, V$, both of which diffuse over time at the diffusion rates $\varepsilon_1$ and $\varepsilon_2$.
In this inverse problem, we aim to infer the parameters $\varepsilon_1$ and $\varepsilon_2$ of this time-dependent PDE based on several observed snapshots of the concentration fields $u$ and $v$.

\begin{figure}[h]
	\centering
	\includegraphics[width=\textwidth]{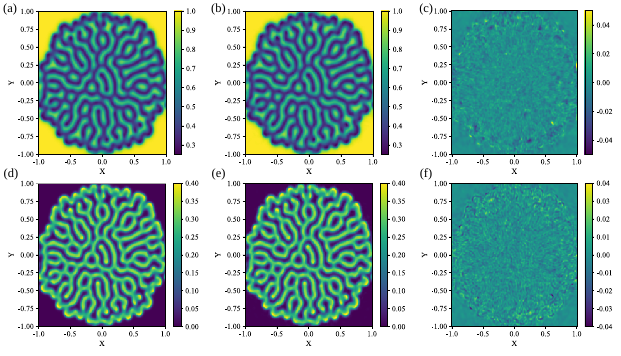}
	\caption{\textbf{Concentration fields of the 2D Gray--Scott model at $\boldsymbol{t=4000}$}. 
		(a) Reference concentration solution $u$.
		(b) Predicted concentration solution $u$.
		(c) Point-wise solution error $\Delta u$.
		(d) Reference concentration solution $v$.
		(e) Predicted concentration solution $v$.
		(f) Point-wise solution error $\Delta v$.
	}
	\label{fig:GSfield}
\end{figure}

Referring to~\cite{Wang2021b}, we generate a dataset containing a numerical solution of the two-dimensional Gray--Scott equations with 40,000 spatial points and 401 temporal snapshots. 
Specifically, we set $b=0.04$ and $d=0.1$, with the reference values of the inverse parameters prescribed as $\varepsilon_1$=2e-5 and $\varepsilon_2$=1e-5. 
Starting from the following initial condition:
\begin{equation}\label{eq:GrayScottBCs}
	\begin{array}{*{20}{rl}}
		u\left(x,y,0 \right)&= 1-\exp\left(-80\left(x+0.05\right)^2+\left(y+0.02\right)^2\right)
		&\left(x,y\right) \in \left[-1,1\right] \times \left[-1,1\right]\\
		v\left(x,y,0 \right)&= \exp\left(-80\left(x-0.05\right)^2+\left(y-0.02\right)^2\right)
		&\left(x,y\right) \in \left[-1,1\right] \times \left[-1,1\right]\\
	\end{array}
\end{equation}
the equations are integrated up to the final time $t=4000$.
Synthetic training data for this example are generated using the Chebfun package~\cite{Chebfun2014} with a spectral Fourier discretization and a fourth-order stiff time-stepping scheme~\cite{Cox2002} with a time-step size of 0.5.
We collect the data points $\left\{\left(t^i,x^i,y^i\right),\left(\bar u^i,\bar v^i\right)\right\}_{i=1}^{N}$ from time $t=3900$ to $t=4000$ (a total of 10 snapshots) as our training dataset.
 Fig.~\ref{fig:GSfield}(a) and (d) display a characteristic fingerprint-like patterns, illustrating a solution with intricate spatial structures and rich frequency components.

To invert the unknown diffusion rates $\varepsilon_1$ and $\varepsilon_2$, we employed the proposed MR-PIKAN model to predict the concentration fields $u$ and $v$, while treating the unknown coefficients $\varepsilon_1, \varepsilon_2$ as trainable parameters.
To better handle the non-trivial frequency features in the concentration fields, we used Chebyshev polynomials of up to the degree 15 as activation functions while retaining the network architecture with two hidden layers, each containing 50 neurons.

The corresponding loss function is given by
\begin{equation}\label{eq:GrayScottLoss}
	\begin{aligned}
		\mathcal L\left({\boldsymbol{\theta}}\right)&=\omega_{ru} \mathcal L_{ru}\left({\boldsymbol{\theta}}\right) + \omega_{rv} \mathcal L_{rv}\left({\boldsymbol{\theta}}\right) + \omega_{du} \mathcal L_{du}\left({\boldsymbol{\theta}}\right) + \omega_{dv} \mathcal L_{dv}\left({\boldsymbol{\theta}}\right)\\
		&=\frac{1}{N_{r}}\sum_{i=1}^{N_{r}}{\left| \mathcal{D}^u\left[u_{\boldsymbol{\theta }} \right] \left( x_r^i,y_r^i,t_r^i \right) \right|^2} + 
		\frac{1}{N_{r}}\sum_{i=1}^{N_{r}}{\left| \mathcal{D}^v\left[v_{\boldsymbol{\theta }} \right] \left( x_r^i,y_r^i,t_r^i \right) \right|^2} \\
		&+\frac{1}{N_{d}}\sum_{i=1}^{N_{d}}{\left| u_{\boldsymbol{\theta }}\left( x_d^i,y_d^i,t_d^i\right) -\bar{u}\left( x_d^i,y_d^i,t_d^i \right) \right|^2}
		+\frac{1}{N_{d}}\sum_{i=1}^{N_{d}}{\left| v_{\boldsymbol{\theta }}\left( x_d^i,y_d^i,t_d^i\right) -\bar{v}\left( x_d^i,y_d^i,t_d^i \right) \right|^2}\\
	\end{aligned}
\end{equation}
where $\left( x_r^i,y_r^i,t_r^i \right)$ and $\left( x_d^i,y_d^i,t_d^i \right)$ represent the residual sampling points and the observed data sampling points, respectively.
$\bar{u}$ and $\bar{v}$ denote the observed values of the concentration fields.
For the single-resolution case, we selected mini-batch sizes of $N_d=1000$ and $N_r=5000$, where all data points were randomly sampled from the observed dataset, and all residual points were randomly sampled within the computational domain.
In particular, since the diffusion rates are strictly positive and generally very small, we parameterized $\varepsilon_1$ and $\varepsilon_2$ using exponential functions, following Wang et al.~\cite{Wang2021b}. Specifically, we set $\varepsilon_i=e^{\alpha_i}$ for $i=1,2$, where $\alpha_1$ and $\alpha_2$ were trainable parameters initialized to $-10$.
All models were trained by minimizing the above loss function via 120,000 iterations of gradient descent. 

\begin{figure}[H]
	\centering
	\includegraphics[width=\textwidth]{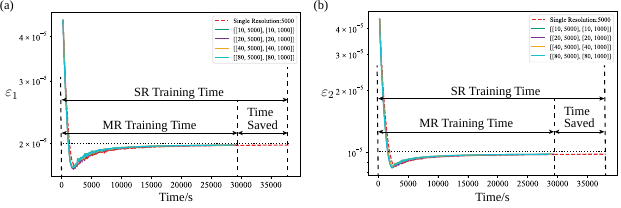}
	\caption{\textbf{Test-error evolution of unknown diffusion coefficients for MR-PIKAN models.}
	(a) Test-error evolution of diffusion coefficient $\varepsilon_1$.
	(b) Test-error evolution of diffusion coefficient $\varepsilon_2$.
	Both the SR-PIKAN and MR-PIKAN models accurately invert the diffusion coefficients, with test errors within 2\%.
	The MR-PIKAN models achieve the same level of accuracy as the SR-PIKAN model while reducing training time by 20\%.
	}
	\label{fig:GS_KAN}
\end{figure}

The predicted solutions and point-wise errors of the concentration fields $u$ and $v$ at $t=4000$ obtained using the MR-PIKAN model are illustrated in Fig.~\ref{fig:GSfield}.
The results show that the predicted concentration fields closely match the reference solution, with only minor discrepancies.
In addition, Fig.~\ref{fig:GS_KAN} illustrates the evolution of the inferred diffusion rates obtained using the SR-PIKAN and MR-PIKAN models.
The final inferred diffusion coefficients are presented in Table~\ref{tab:varepsilon}, showing that the coefficients inferred by both SR-PIKAN and MR-PIKAN are in good agreement with the exact values.
The inversion errors of all PIKAN models are within 2\%, while the MR-PIKAN models reduce the training time by 20\% compared to the SR-PIKAN model.
Overall, the proposed MR-PIKAN framework achieves accurate field reconstruction and parameter identification for this inverse problem, while improving training efficiency relative to SR-PIKAN.
\renewcommand{\arraystretch}{1.25}
    \begin{table}[H]
	\footnotesize
	\centering
	\begin{tabular}{ccccccc}
		\hline
        \multirow{3}{*}{Model} &\multirow{3}{*}{Resolution} &\multicolumn{2}{c}{$\varepsilon_1$} &\multicolumn{2}{c}{$\varepsilon_2$} &\multirow{3}{*}{Training Time (s)}\\
        \cline{3-6}
        & &Inferred&Relative&Inferred&Relative &\\
        & &Value &Error &Value &Error &\\
        \hline
        SR-PIKAN&5000 &1.977e-5 &1.15\% &9.832e-6 &1.68\% &38026\\
        \hline
        \multirow{4}{*}{MR-PIKAN}&$\left[10,5000\right],\left[10,1000\right]$ &1.977e-5 &1.15\% &9.830e-6 &1.70\% &29299\\
		\cline{2-7}
        &$\left[20,5000\right],\left[20,1000\right]$ &1.976e-5 &1.20\% &9.912e-6 &1.88\% &29347\\
		\cline{2-7}
        &$\left[40,5000\right],\left[40,1000\right]$ &1.979e-5 &1.05\% &9.833e-6 &1.67\% &28959\\
		\cline{2-7}
        &$\left[80,5000\right],\left[80,1000\right]$ &1.978e-5 &1.10\% &9.834e-6 &1.66\% &28554\\
		\hline
		\end{tabular}
	\caption{Comparison of the inversion errors and training times of the SR-PIKAN and MR-PIKAN models for the inverse 2D Gray--Scott problem. (Exact parameter values: $\varepsilon_{1}$=2e-5, $\varepsilon_{2}$=1e-5)}
    \label{tab:varepsilon}
    \end{table}
\renewcommand{\arraystretch}{1}

\subsection{Hidden elasticity of two-phase random materials}
\begin{figure}[h]
	\centering
	\includegraphics[width=0.45\textwidth]{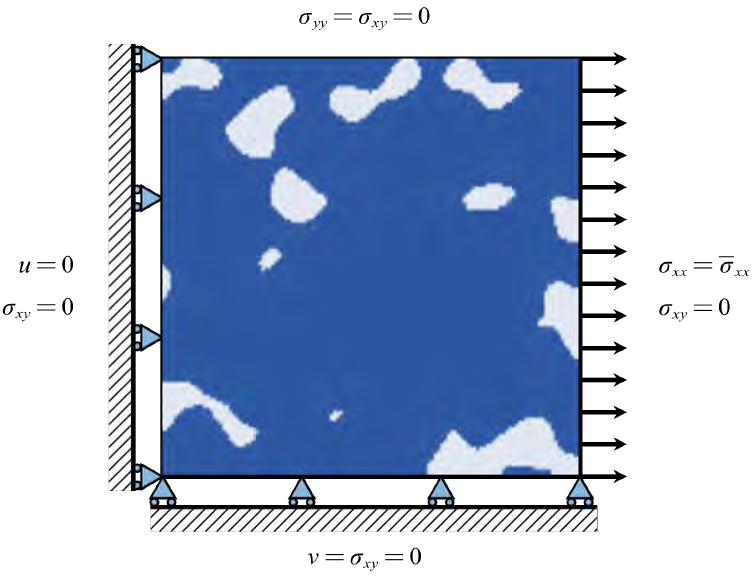}
	\caption{Microstructure of TRMs and boundary conditions of the problem.}
	\label{fig:TRMs}
\end{figure}

The last example further evaluates the performance of the MR-PIKAN model in an inverse problem involving the reconstruction of parameter fields with complex multi-scale features.
Two-phase random materials (TRMs), such as polyurethane~\cite{Liu2024}, multi-phase metallic alloys~\cite{Eckermann2014} and nanocomposites~\cite{Iyer2019}, are an important class of materials in advanced material research whose properties are closely related to their microstructures.
In this example, we consider a two-phase square plate with a length of 2 mm.
The microstructure of this TRM plate, as shown in  Fig.~\ref{fig:TRMs}, is constructed using the random field level-cutting method~\cite{Ren2024}.

Assuming that the interfaces between the two phases are perfectly bonded and each phase is an isotropic elastic material, the governing equations can be expressed in the following form:
\begin{equation}\label{linearmomentum}
	\begin{aligned}
	\sigma_{ij,j}&=0\\
	\sigma_{ij}	&=\lambda \delta_{ij}\varepsilon_{kk}+2\mu \varepsilon_{ij}\\
	\varepsilon_{ij}&=\frac{1}{2}\left(u_{i,j}+u_{j,i}\right)
	\end{aligned}
\end{equation}
where $\sigma_{ij}$ and $\varepsilon_{ij}$ denote the components of the Cauchy stress tensor and the infinitesimal strain tensor, respectively, and $\varepsilon_{ij}$ is derived from the displacement components $u_i$.
The material parameter $\lambda$ and $\mu$, known as Lam\'{e} constants, can be related to the elastic modulus and Poisson's ratio through the following equations:
\begin{equation}\label{lame}
	\begin{aligned}
		E&=\frac{\mu \left(3\lambda+2\mu\right)}{\left(\lambda+\mu\right)}\\
		\nu	&= \frac{\lambda}{2\left(\lambda+\mu\right)}
	\end{aligned}
\end{equation}
In practical experimental settings, for a given two-phase material plate subjected to known surface force boundary conditions, global strain-field data can be obtained non-destructively through various imaging techniques~\cite{DIC2015,Chen2021}.
Therefore, in this inverse problem, we aim to infer the material parameter fields $E\left(x,y \right)$ and $\nu \left(x,y \right)$ of this two-phase material plate based on observed strain fields and the known surface force boundary conditions.

\begin{figure}[H]
	\centering
	\includegraphics[width=\textwidth]{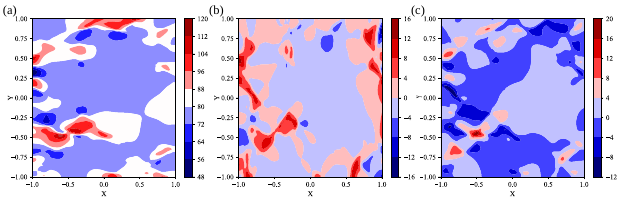}
	\caption{\textbf{Reference stress fields of TRM plates.}
	(a) Stress field $\sigma_{xx}$.
	(b) Stress field $\sigma_{yy}$.
	(c) Stress field $\sigma_{xy}$.
	The reference stress components $\left\{\left( \sigma_{xx}^i,\sigma_{yy}^i\right)\right\}_{i=1}^{N_\sigma}$ at the boundaries are consistent with the known surface force boundary conditions $\left\{\left(\bar \sigma_{xx}^i,\bar\sigma_{yy}^i\right)\right\}_{i=1}^{N_\sigma}$.
    Together with the strain-field data $\left\{\left(\bar\varepsilon_{xx}^i,\bar\varepsilon_{yy}^i,\bar\varepsilon_{xy}^i\right)\right\}_{i=1}^{N_\varepsilon}$ within the domain, these constitute the observed training dataset.
	}
	\label{fig:TRM_sigma}
\end{figure}

\begin{figure}[H]
	\centering
	\includegraphics[width=\textwidth]{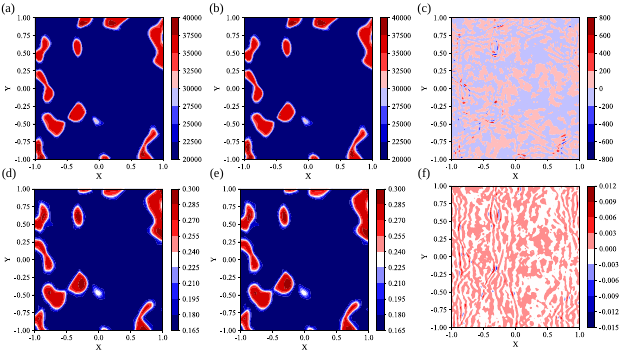}
	\caption{\textbf{Material parameter field results of the TRM plate}. 
		(a) Reference Young's modulus $E$.
		(b) Predicted Young's modulus $E$.
		(c) Point-wise solution error of Young's modulus $\Delta E$.
		(d) Reference Poisson's ratio $\nu$.
		(e) Predicted Poisson's ratio $\nu$.
		(f) Point-wise solution error of Poisson's ratio $\Delta \nu$.
	}
	\label{fig:TRM_field}
\end{figure}

The FEM method was used to generate the numerical solution with $128\times128$ elements.
The boundary conditions are illustrated in  Fig.~\ref{fig:TRMs}, where the right side of the plate is subjected to a surface traction of $\bar{\sigma}_{xx}=80$MPa.
 Fig.~\ref{fig:TRM_field}(a) and (d) show the exact elastic modulus field and Poisson's ratio field of the plate, respectively, which are the target parameter fields to be identified in this example.
We collect the observed strain components $\left\{\left(\bar \varepsilon_{xx}^i,\bar \varepsilon_{yy}^i,\bar \varepsilon_{xy}^i\right)\right\}_{i=1}^{N_\varepsilon}$ and the stress components $\left\{\left(\bar \sigma_{xx}^i,\bar \sigma_{yy}^i\right)\right\}_{i=1}^{N_\sigma}$ on the boundaries as the training data set.
The stress fields obtained from the reference FEM solution are shown in  Fig.~\ref{fig:TRM_sigma}, where pronounced stress concentration is observed.

To invert the parameter fields of this two-phase material, we utilized the proposed MR-PIKAN framework to predict the stress fields and material parameter fields of the aforementioned plate.
To better handle the non-trivial frequency features of two-phase materials, we employed Chebyshev polynomials of up to degree 15 as activation functions, while maintaining a network architecture with three hidden layers, each consisting of 60 neurons.
To effectively represent the stress fields and material parameter fields with multiple components using a single neural network, we modified the final layer of the standard KAN structure into a learnable linear layer, thereby accelerating the model's convergence.
To address the challenge of learning two types of field variables with significant amplitude differences, we introduced scaling factors and derived their equivalent dimensionless governing equations, following the standard procedure described in~\cite{Langtangen2016}.
The corresponding loss function is given by:
\begin{equation}\label{eq:TRMLosses}
	\begin{aligned}
		\mathcal L\left({\boldsymbol{\theta}}\right)&=\omega_{\sigma_{xx}} \mathcal L_{\sigma_{xx}}
		+\omega_{\sigma_{yy}} \mathcal L_{\sigma_{yy}}
		+\omega_{cxx} \mathcal L_{cxx}
		+\omega_{cyy} \mathcal L_{cyy}
		+\omega_{cxy} \mathcal L_{cxy}
		+\omega_{rx} \mathcal L_{rx}
		+\omega_{ry} \mathcal L_{ry}\\
		&=\frac{\omega_{\sigma_{xx}}}{N_{\sigma}}\sum_{i=1}^{N_{\sigma}}{\left|\sigma_{xx}^{\boldsymbol{\theta }}-\bar{\sigma}_{xx}\right|^2}+
		\frac{\omega_{\sigma_{yy}}}{N_{\sigma}}\sum_{i=1}^{N_{\sigma}}{\left|\sigma_{yy}^{\boldsymbol{\theta }}-\bar{\sigma}_{yy}\right|^2}
		+\frac{\omega_{cxy}}{N_{c}}\sum_{i=1}^{N_{c}}{\left|2\mu \bar \varepsilon_{xy}-\sigma_{xy}^{\boldsymbol{\theta }}\right|^2}\\
		&+\frac{\omega_{cxx}}{N_{c}}\sum_{i=1}^{N_{c}}{\left|\left(\lambda+2\mu\right)\bar \varepsilon_{xx}+\lambda\bar\varepsilon_{yy}-\sigma_{xx}^{\boldsymbol{\theta }}\right|^2}
		+\frac{\omega_{cyy}}{N_{c}}\sum_{i=1}^{N_{c}}{\left|\left(\lambda+2\mu\right)\bar\varepsilon_{yy}+\lambda\bar\varepsilon_{xx}-\sigma_{yy}^{\boldsymbol{\theta }}\right|^2}\\
		&+\frac{\omega_{rx}}{N_{r}}\sum_{i=1}^{N_{r}}{\left|\sigma_{xx,x}^{\boldsymbol{\theta }}+\sigma_{xy,y}^{\boldsymbol{\theta }}\right|^2}+
		\frac{\omega_{ry}}{N_{r}}\sum_{i=1}^{N_{r}}{\left|\sigma_{yx,x}^{\boldsymbol{\theta }}+\sigma_{yy,y}^{\boldsymbol{\theta }}\right|^2}		
	\end{aligned}
\end{equation}
where variables with the superscript $\boldsymbol{\theta}$ represent the results approximated by the neural network.
$\bar\varepsilon_{xx}, \bar\varepsilon_{yy}, \bar\varepsilon_{xy}$ denote the observed strain components, and $\bar\sigma_{xx}, \bar\sigma_{yy}$ denote the prescribed stress components on the boundary.
To avoid training pathologies and better balance the seven loss terms, we employed the adaptive weighting method proposed by Wang et al. in their work~\cite{Wang2021a}.
All models were trained for 400,000 iterations, and the initial learning rate was set to 1e-4.
For the single-resolution case, we chose the mini-batch size $N_{\sigma}=N_c=N_r=2000$. 
Moreover, we trained the same PIKAN model with an alternating multi-resolution strategy.

\begin{figure}[h]
	\centering
	\includegraphics[width=\textwidth]{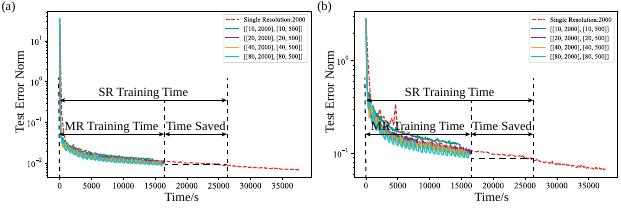}
	\caption{\textbf{Test-error evolution of SR-PIKAN and MR-PIKAN models achieving comparable error norms.}
		(a) Test-error evolution of material parameter fields.
		(b) Test-error evolution of stress fields.
		It can be observed that the MR-PIKAN models significantly reduce the training time required to achieve acceptable training accuracy compared to the SR-PIKAN model.
	}
	\label{fig:TRM_KAN}
\end{figure}

The predicted solution and point-wise error of the material property fields learned by the MR-PIKAN model are shown in  Fig.~\ref{fig:TRM_field}.
The results show a strong alignment between the predictions and the ground truth of the parameter fields. Fig.~\ref{fig:TRM_KAN} shows the evolution of the $L_2$ error of the models trained using the multi-resolution training strategy with different step sizes.
The results show that the proposed MR-PIKAN model requires less than half the training time of the single high-resolution model for the same number of iterations.
Moreover, the time taken by the MR-PIKAN model to achieve an acceptable accuracy (within 1\%) for the inferred material parameter fields is over 30\% less than that required by the SR-PIKAN model.
These results indicate that the proposed MR-PIKAN model can accurately infer material property fields with sharp discontinuities, while substantially reducing the computational cost compared with the SR-PIKAN model.

\section{Conclusions}
\label{sec:Conclusions}
PDE problems with multi-scale features remain challenging for physics-informed learning methods for two main reasons. First, conventional neural-network representations often suffer from spectral bias, which limits their ability to resolve high-frequency or localized solution components. Second, accurately capturing such multi-scale features usually requires dense residual sampling, which substantially increases the computational cost. In this work, we propose MR-PIKAN, a multi-resolution training-enhanced PIKAN framework, which trains the model sequentially or alternately at different resolutions. The key idea is to combine the representation capability of KANs with multi-resolution sampling. Training at lower resolutions reduces the cost of gradient evaluation, whereas training at higher resolutions provides richer residual information for resolving multi-scale solution features. This design allows the training process to balance computational efficiency and solution accuracy.

The numerical results demonstrate that MR-PIKAN achieves accuracy comparable to high-resolution training while substantially reducing training time. In the forward multi-scale benchmark, the alternating multi-resolution strategy achieved faster convergence than the sequential strategy. These findings are consistent with the NTK-based eigenvalue spectral analysis, which suggests that the KAN architecture has advantages in capturing multi-scale solution components. The training-dynamics analysis based on information bottleneck theory further shows that PIKAN models exhibits behavior similar to Fourier-feature PINN models when resolving multi-scale features, whereas vanilla PINN follows a different training pattern associated with its limited ability to capture high-frequency components. The applicability of MR-PIKAN was further examined through inverse problems involving multi-scale fields and discontinuous material properties. In the Gray--Scott inverse problem, MR-PIKAN accurately reconstructed concentration fields and identified diffusion coefficients with small inversion errors. In the heterogeneous material example, MR-PIKAN successfully recovered discontinuous material parameter fields, suggesting its potential for inverse modelling of multi-phase materials.

Despite these advantages, the present study still has several limitations. The resolution schedule, including the choice of low and high resolutions and the switching interval, is currently prescribed empirically, which may limit the robustness of the method across different multi-scale PDE problems. Future work will focus on developing adaptive resolution-switching criteria based on residuals, error indicators, or training dynamics, rather than relying on fixed schedules. In addition, extending the framework to noisy experimental inverse problems remains an important direction, where measurement uncertainty and data incompleteness may further challenge the stability of parameter identification. Addressing these issues may further improve the robustness and applicability of MR-PIKAN as an efficient physics-informed learning framework for complex multi-scale PDE problems.

\section*{CRediT authorship contribution statement}
\noindent
\textbf{Yu-Sen Yang}: Conceptualization, Investigation, Methodology, Software, Visualization, Writing -- original draft, Writing -- review \& editing;
\textbf{Ling Guo}: Conceptualization, Funding acquisition, Investigation, Methodology, Supervision, Writing -- review \& editing; 
\textbf{Xiaodan Ren}: Conceptualization, Funding acquisition, Investigation, Supervision, Writing -- review \& editing; 

\section*{Acknowledgements}
This work was supported by the National Natural Science Foundation of China (under grant numbers 92270115, 12071301) to Ling Guo; the National Natural Science Foundation of China (Grant No. 52478198) and Shanghai Municipal Science and Technology Commission (Grant No. 24510711500) to Xiaodan Ren.

\appendix

\section{Implementation details of PIKAN and PINN models}
\label{sec:appendixA}
The network parameters are initialized using the Xavier normal distribution~\cite{Xavier2010}. The rest of the case-specific information is summarized in Table.~\ref{tab:implementationPoisson}-Table.~\ref{tab:implementationTRM}.

\renewcommand{\arraystretch}{1.25}
\begin{table}[h]
	\footnotesize
	\centering
	\begin{tabularx}{\textwidth}{cl>{\centering\arraybackslash}X>{\centering\arraybackslash}X}
		\toprule
		\multicolumn{4}{l}{Implementation details: forward 2D Poisson problem}\\
		\midrule
		\multicolumn{2}{l}{Network approximation} & \multicolumn{2}{c}{$u_{\boldsymbol{\theta}}\left(x,y\right)$}\\
		\multirow{6}{*}{Network details} &Architecture & KAN &MLP\\
		&Number of hidden layers &2 &6\\
		&Neurons per layer &60 &72\\
		&\multirow{2}{*}{Activation function} &Chebyshev polynomials &\multirow{2}{*}{$\tanh$}\\
		& &$T_0\left( x \right) \cdots T_6\left( x \right)$ &\\
		&Number of parameters &26700 &27433\\
		\multicolumn{2}{l}{Initial learning rate} & \multicolumn{2}{c}{1e-3} \\
		\multicolumn{2}{l}{Full-batch size of collocation points} & SR:5120 &MR:1280, 2560, 5120\\
		\multicolumn{2}{l}{Full-batch size of boundary condition points} & \multicolumn{2}{c}{800} \\
		\multicolumn{2}{l}{Number of epochs} & \multicolumn{2}{c}{40000} \\
		\multicolumn{2}{l}{Decay rate} & \multicolumn{2}{c}{0.9} \\
		\multicolumn{2}{l}{Decay step} & \multicolumn{2}{c}{5000} \\
		\multicolumn{2}{l}{Global weight of residual term} & \multicolumn{2}{c}{0.1} \\
		\multicolumn{2}{l}{Global weight of boundary condition term} & \multicolumn{2}{c}{100} \\
		\bottomrule
	\end{tabularx}
	\caption{Implementation details of the forward Poisson case.}
	\label{tab:implementationPoisson}
\end{table}
\renewcommand{\arraystretch}{1}

\renewcommand{\arraystretch}{1.25}
\begin{table}[H]
	\footnotesize
	\centering
	\begin{tabularx}{\textwidth}{l>{\centering\arraybackslash}X>{\centering\arraybackslash}X}
		\toprule
		\multicolumn{3}{l}{Implementation details: inverse Gray--Scott model} \\
		\midrule
		Network approximation &\multicolumn{2}{c}{$\left[ u_{\boldsymbol{\theta}}\left(x,y,t\right), v_{\boldsymbol{\theta}}\left(x,y,t\right) \right]$} \\
		Network architecture &\multicolumn{2}{c}{KAN} \\
		Number of hidden layers &\multicolumn{2}{c}{2}\\
		Neurons per layer &\multicolumn{2}{c}{50}\\
		Activation function &\multicolumn{2}{c}{Chebyshev polynomials: $T_0\left( x \right) \cdots T_{15}\left( x \right)$}\\
		Number of parameters &\multicolumn{2}{c}{44200}\\
		Initial learning rate &\multicolumn{2}{c}{1e-3} \\
		Number of data points &\multicolumn{2}{c}{400000}\\
            Mini-batch size of collocation points & SR: 5000 & MR: 1000, 5000 \\
		Mini-batch size of data points & SR: 1000 & MR: 1000, 1000\\
		Number of epochs & \multicolumn{2}{c}{120000} \\
		Decay rate & \multicolumn{2}{c}{0.9} \\
		Decay step & \multicolumn{2}{c}{5000} \\
		Global weight of each loss term & \multicolumn{2}{c}{1.0} \\
		\bottomrule
	\end{tabularx}
	\caption{Implementation details of the inverse Gray--Scott model.}
	\label{tab:implementationGray-Scott}
\end{table}
\renewcommand{\arraystretch}{1}

\renewcommand{\arraystretch}{1.25}
\begin{table}[H]
	\footnotesize
	\centering
	\begin{tabularx}{\textwidth}{l>{\centering\arraybackslash}X>{\centering\arraybackslash}X}
		\toprule
		\multicolumn{3}{l}{Implementation details: hidden elasticity of TRMs} \\
		\midrule
		Network approximation &\multicolumn{2}{c}{$\left[\sigma_{xx}^{\boldsymbol{\theta}}(x,y), \sigma_{yy}^{\boldsymbol{\theta}}(x,y), \sigma_{xy}^{\boldsymbol{\theta}}(x,y)\right]$ and $\left[\lambda^{\boldsymbol{\theta}}(x,y), \mu^{\boldsymbol{\theta}}(x,y)\right]$} \\
		Network architecture & \multicolumn{2}{c}{KAN} \\
		Number of hidden KAN layers & \multicolumn{2}{c}{3} \\
		Neurons per layer & \multicolumn{2}{c}{60} \\
		Activation function & \multicolumn{2}{c}{Chebyshev polynomials: $T_0(x) \cdots T_{15}(x)$} \\
		Number of parameters & \multicolumn{2}{c}{117303 and 117242} \\
		Initial learning rate & \multicolumn{2}{c}{1e-4} \\
            Number of data points & \multicolumn{2}{c}{16384} \\
		Mini-batch size of collocation points & SR: 2000 & MR: 500, 2000 \\
		Mini-batch size of data points & SR: 2000 & MR: 500, 2000 \\
		Number of epochs & \multicolumn{2}{c}{4000} \\
		Number of iterations per epoch & \multicolumn{2}{c}{100} \\
		Decay rate & \multicolumn{2}{c}{0.9} \\
		Decay step & \multicolumn{2}{c}{5000} \\
		Initial global weight of each loss term & \multicolumn{2}{c}{1.0} \\
		\bottomrule
	\end{tabularx}
	\caption{Implementation details of the hidden elasticity case.}
	\label{tab:implementationTRM}
\end{table}
\renewcommand{\arraystretch}{1}

\section{Training dynamics of PINNs from the NTK perspective}
\label{sec:appendixNTK}
	According to the Neural Tangent Kernel (NTK) theory~\cite{NEURIPS2018}, the training dynamics of a sufficiently wide neural network under gradient flow can be approximately characterized by a kernel that remains nearly constant during training. For PINNs, the NTK framework can be further extended to characterize the coupled evolution of the boundary error and PDE residual error~\cite{wangWhenWhyPINNs2022}. For the Poisson problem considered in Section~\ref{subsec:SR}, the loss function defined in Eq.~\eqref{eq:2DPoissonLoss} induces the following continuous-time gradient-flow dynamics:
	\begin{equation}
		\begin{aligned}
			\frac{\mathrm{d}}{\mathrm{d}t}
			\begin{bmatrix}
				u_{\boldsymbol{\theta}}(\boldsymbol{x}_b)-g\left(\boldsymbol{x}_b\right) \\[2mm]
				\Delta u_{\boldsymbol{\theta}}\left(\boldsymbol{x}_r\right)+f\left(\boldsymbol{x}_r\right)
			\end{bmatrix}
			&=
			-
			\begin{bmatrix}
				\frac{\omega_b}{N_b}\mathbf{K}_{uu}(t) 
				&
				\frac{\omega_r}{N_r}\mathbf{K}_{ur}(t)
				\\[2mm]
				\frac{\omega_b}{N_b}\mathbf{K}_{ru}(t) 
				&
				\frac{\omega_r}{N_r}\mathbf{K}_{rr}(t)
			\end{bmatrix}
			\begin{bmatrix}
				u_{\boldsymbol{\theta}}(\boldsymbol{x}_b)-g\left(\boldsymbol{x}_b\right) \\[2mm]
				\Delta u_{\boldsymbol{\theta}}\left(\boldsymbol{x}_r\right)+f\left(\boldsymbol{x}_r\right)
			\end{bmatrix} \\
			&:=
			-\mathbf{K}(t)
			\begin{bmatrix}
				u_{\boldsymbol{\theta}}(\boldsymbol{x}_b)-g\left(\boldsymbol{x}_b\right) \\[2mm]
				\Delta u_{\boldsymbol{\theta}}\left(\boldsymbol{x}_r\right)+f\left(\boldsymbol{x}_r\right)
			\end{bmatrix}.
		\end{aligned}
	\end{equation}
	For clarity, we denote the training error vector by
	\begin{equation}
		\boldsymbol e(t)
		=
		\begin{bmatrix}
			u_{\boldsymbol{\theta}}(\boldsymbol{x}_b)-g\left(\boldsymbol{x}_b\right) \\[2mm]
			\Delta u_{\boldsymbol{\theta}}\left(\boldsymbol{x}_r\right)+f\left(\boldsymbol{x}_r\right)
		\end{bmatrix}.
	\end{equation}
	Following the frozen-NTK approximation used in~\cite{wangWhenWhyPINNs2022}, the NTK matrix is assumed to vary slowly during training and can be approximated by its initial value, $\mathbf{K}(t)\approx \mathbf{K}(0)$, and we denote the frozen NTK dynamics matrix by $\mathbf{K}_{\mathrm{NTK}}:=\mathbf{K}(0)$. The above dynamics therefore admit the following approximate solution:
	\begin{equation}
		\boldsymbol e(t)
		\approx
		\exp\left(-\mathbf{K}_{\mathrm{NTK}}t\right)
		\boldsymbol e(0),
	\end{equation}
	where $\boldsymbol e(0)$ denotes the initial coupled error vector.
	
	To interpret the convergence behavior from the NTK spectrum, we consider the eigendecomposition of the frozen NTK dynamics matrix. If $\mathbf{K}_{\mathrm{NTK}}$ is diagonalizable as
	\begin{equation}
		\mathbf{K}_{\mathrm{NTK}}
		=
		\mathbf{V}\boldsymbol{\Lambda}\mathbf{V}^{-1},
	\end{equation}
	where $\mathbf{V}$ is the matrix whose columns are the eigenvectors of $\mathbf{K}_{\mathrm{NTK}}$, and $\boldsymbol{\Lambda}$ is a diagonal matrix containing the corresponding eigenvalues, then the error evolution can be written as
	\begin{equation}
		\mathbf{V}^{-1}\boldsymbol e(t)
		\approx
		\exp\left(-\boldsymbol{\Lambda}t\right)
		\mathbf{V}^{-1}\boldsymbol e(0).
	\end{equation}
	This expression shows that different error modes decay at rates governed by the eigenvalues of $\mathbf{K}_{\mathrm{NTK}}$. Larger eigenvalues correspond to modes with faster convergence, whereas smaller eigenvalues indicate components that are learned more slowly. Therefore, the NTK eigenvalue distribution provides a perspective on the training dynamics for comparing the trainability and spectral bias behavior of PINN and PIKAN models.

\section{Quantitative assessment of computational efficiency}
\label{app:metrics}

To quantitatively evaluate the computational efficiency of different methods, we report the number of training iterations, total wall-clock time, average wall-clock time per iteration, and peak GPU memory consumption for each numerical example. As summarized in Table~\ref{tab:metric}, these metrics provide a direct comparison of the training cost among the models considered in each example. In this table, MR-PIKAN refers to the alternating multi-resolution training implementation with an alternating interval of 20 iterations. The average wall-clock time per iteration is computed as the total wall-clock time divided by the number of training iterations. All experiments were conducted under the same hardware and software settings to ensure a fair comparison.

\renewcommand{\arraystretch}{1.25}
\begin{table}[htbp]
	\centering
	\footnotesize
	\begin{tabular}{llcccc}
		\toprule
		\makecell{Example} 
		& \makecell{Method} 
		& \makecell{Iterations} 
		& \makecell{Wall-clock\\time (s)} 
		& \makecell{Time per\\iteration (s)}
		& \makecell{Peak GPU\\memory (MB)} \\
		\midrule
		\multirow{3}{*}{\makecell{Forward Poisson problem}}
		& SR-PINN   & 40000 & 2595 & 6.488e-02 & 1164.0\\
		& SR-PIKAN  & 40000 & 2513 & 6.283e-02 & 7386.9\\
		& MR-PIKAN  & 40000 & 2142 & 5.355e-02 & 7386.9\\
		
		\midrule
		\multirow{2}{*}{\makecell{Gray--Scott inverse problem}}
		& SR-PIKAN  & 120000 & 38026 & 3.169e-01 & 5888.8\\
		& MR-PIKAN  & 120000 & 29347 & 2.446e-01 & 5888.8\\
		
		\midrule
		\multirow{2}{*}{\makecell{Hidden elasticity in TRMs}}
		& SR-PIKAN  & 400000 & 37861 & 9.465e-02 & 1042.3\\
		& MR-PIKAN  & 400000 & 16625 & 4.156e-02 & 1042.3\\
		
		\bottomrule
	\end{tabular}
	\caption{Quantitative assessment of computational efficiency for different models across numerical examples.}
	\label{tab:metric}
\end{table}
\renewcommand{\arraystretch}{1}


 \bibliographystyle{elsarticle}
 \bibliography{refs}

@article{singhDeepMachineLearning2024,
	title = {Deep Machine Learning Approaches for Battery Health Monitoring},
	author = {Singh, S. and Budarapu, P. R.},
	year = 2024,
	month = aug,
	journal = {Energy},
	volume = {300},
	pages = {131540},
	issn = {0360-5442},
	doi = {10.1016/j.energy.2024.131540},
	urldate = {2026-05-23},
}

@article{deshpandePerformanceLifeAnalysis2025,
	title = {Performance and {{Life Analysis}} of {{Lithium-Ion Batteries Aided}} by {{Data-Driven Analysis}}},
	author = {Deshpande, Ashish Anil and Siruvuri, S. D. V. S. S. Varma and Sastry, Y. B. Sudhir and Rammohan, Bhanumurthy and Mahmoud, Samy Refahy and Budarapu, Pattabhi Ramaiah},
	year = 2025,
	journal = {Int. J. Mech. Syst. Dyn.},
	volume = {5},
	number = {2},
	pages = {277--289},
	issn = {2767-1402},
	doi = {10.1002/msd2.70014},
	urldate = {2026-06-10},
	copyright = {\copyright{} 2025 The Author(s). International Journal of Mechanical System Dynamics published by John Wiley \& Sons Australia, Ltd on behalf of Nanjing University of Science and Technology.},
	langid = {english}
}

@article{raissiPhysicsinformedNeuralNetworks2019,
	title = {Physics-Informed Neural Networks: {{A}} Deep Learning Framework for Solving Forward and Inverse Problems Involving Nonlinear Partial Differential Equations},
	shorttitle = {Physics-Informed Neural Networks},
	author = {Raissi, M. and Perdikaris, P. and Karniadakis, G.E.},
	year = 2019,
	month = feb,
	journal = {J. Comput. Phys.},
	volume = {378},
	pages = {686--707},
	issn = {00219991},
	doi = {10.1016/j.jcp.2018.10.045},
	urldate = {2025-02-02},
	langid = {english},
}

@article{Karniadakis2021,
	title={Physics-informed machine learning},
	author={Karniadakis, George Em and Kevrekidis, Ioannis G and Lu, Lu and Perdikaris, Paris and Wang, Sifan and Yang, Liu},
	journal={Nat. Rev. Phys.},
	volume={3},
	number={6},
	pages={422--440},
	year={2021},
	doi = {10.1038/s42254-021-00314-5},
	publisher={Nature Publishing Group UK London}
}

@article{Zhang2025,
	title = {Inverse Problems Solutions using Physics-Informed Neural Networks with Novel Adaptive Activation Function},
	journal = {Appl. Math. Model.},
	pages = {116604},
	year = {2025},
	volume = {152},
	issn = {0307-904X},
	doi = {10.1016/j.apm.2025.116604},
	author = {Jun Zhang and Timon Rabczuk and Chensen Ding},
}

@article{Wu2025,
	title = {A physics-informed machine learning framework for inverse multi-solution problems},
	journal = {Appl. Math. Model.},
	pages = {116618},
	year = {2025},
	volume = {152},
	issn = {0307-904X},
	doi = {10.1016/j.apm.2025.116618},
	author = {Zexiong Wu and Xueyou Li and Jian-Hong Wan and Tianhua Xu},
}

@article{KAN2024b,
	author = {Liu, Ziming and Ma, Pingchuan and Wang, Yixuan and Matusik, Wojciech and Tegmark, Max},
	title = {Kan 2.0: Kolmogorov-arnold networks meet science},
	journal = {arXiv preprint arXiv:2408.10205},
	year = {2024},
	doi = {10.48550/arXiv.2408.10205},
	type = {Journal Article}
}

@inproceedings{KAN2024a,
	title = {{{KAN}}: {{Kolmogorov}}--Arnold Networks},
	booktitle = {International Conference on Learning Representations},
	author = {Liu, Ziming and Wang, Yixuan and Vaidya, Sachin and Ruehle, Fabian and Halverson, James and Soljacic, Marin and Hou, Thomas and Tegmark, Max},
	editor = {Yue, Y. and Garg, A. and Peng, N. and Sha, F. and Yu, R.},
	year = 2025,
	volume = {2025},
	pages = {70367--70413}
}

@article{FAIRKAN2024,
    title = {A comprehensive and FAIR comparison between MLP and KAN representations for differential equations and operator networks},
    journal = {Comput. Methods Appl. Mech. Eng.},
    volume = {431},
    pages = {117290},
    year = {2024},
    issn = {0045-7825},
    doi = {10.1016/j.cma.2024.117290},
    author = {Khemraj Shukla and Juan Diego Toscano and Zhicheng Wang and Zongren Zou and George Em Karniadakis}
}

@article{WaveletKAN,
	author = {Bozorgasl, Zavareh and Chen, Hao},
	title = {Wav-kan: Wavelet kolmogorov-arnold networks},
	journal = {arXiv preprint arXiv:2405.12832},
	year = {2024},
	doi = {10.48550/arXiv.2405.12832},
	type = {Journal Article}
}

@inproceedings{Lin2023,
    title={Accelerating the Training of Video Super-resolution Models}, 
    volume={37(2)},
    booktitle = {Proceedings of the AAAI Conference on Artificial Intelligence}, 
    author={Lin, Lijian and Wang, Xintao and Qi, Zhongang and Shan, Ying}, 
    year={2023}, 
    month={June},
    doi = {10.1609/AAAI.V37I2.25246}, 
    pages={1595-1603} 
}

@inproceedings{Facebook2020,
    author = {Wu, Chao-Yuan and Girshick, Ross and He, Kaiming and Feichtenhofer, Christoph and Krahenbuhl, Philipp},
    title = {A Multigrid Method for Efficiently Training Video Models},
    booktitle = {Proceedings of the IEEE/CVF Conference on Computer Vision and Pattern Recognition (CVPR)},
    month = {June},
    year = {2020},
    doi = {10.1109/CVPR42600.2020.00023},
    pages={153-162}
}

@article{RBFKAN,
	author = {Li, Ziyao},
	title = {Kolmogorov-arnold networks are radial basis function networks},
	journal = {arXiv preprint arXiv:2405.06721},
	year = {2024},
	doi = {10.48550/arXiv.2405.06721},
	type = {Journal Article}
}

@article{ChebyKAN,
	author = {Sidharth, SS and Keerthana, AR and Gokul, R and Anas, KP},
	title = {Chebyshev Polynomial-Based Kolmogorov-Arnold Networks: An Efficient Architecture for Nonlinear Function Approximation},
	journal = {arXiv preprint arXiv:2405.07200},
	year = {2024},
	doi = {10.48550/arXiv.2405.07200},
	type = {Journal Article}
}

@book{Hackbusch,
  author    = {Wolfgang Hackbusch},
  title     = {Multi-Grid Methods and Applications},
  publisher = {Springer Berlin, Heidelberg},
  year      = {1985},
  series    = {Springer Series in Computational Mathematics},
  doi = {10.1007/978-3-662-02427-0},
  volume    = {4},
}

@article{E2018,
	title={The Deep Ritz Method: A Deep Learning-Based Numerical Algorithm for Solving Variational Problems},
	author={Weinan, E and Yu, Bing},
	journal={Commun. Math. Stat.},
	volume={6},
	number={1},
	pages={1-12},
	year={2018},
	doi = {10.1007/s40304-018-0127-z},
	publisher={Springer Verlag}
}

@inproceedings{Kolmogorov1957,
    title={On the representation of continuous functions of many variables by superposition of continuous functions of one variable and addition},
    author={Kolmogorov, Andrei Nikolaevich},
    booktitle={Doklady Akademii Nauk SSSR},
    volume={114(5)},
    pages={953-956},
    year={1957},
    organization={Russian Academy of Sciences},
}

@article{Samaniego2020,
	title={An energy approach to the solution of partial differential equations in computational mechanics via machine learning: Concepts, implementation and applications},
	author={Samaniego, Esteban and Anitescu, Cosmin and Goswami, Somdatta and Nguyen-Thanh, Vien Minh and Guo, Hongwei and Hamdia, Khader and Zhuang, X and Rabczuk, T},
	journal={Comput. Methods Appl. Mech. Eng.},
	volume={362},
	pages={112790},
	year={2020},
	doi = {10.1016/j.cma.2019.112790},
	publisher={Elsevier}
}

@inproceedings{Xavier2010,
  title ={Understanding the difficulty of training deep feedforward neural networks},
  author = {Glorot, Xavier and Bengio, Yoshua},
  booktitle = {Proceedings of the Thirteenth International Conference on Artificial Intelligence and Statistics},
  pages = {249-256},
  year = 	{2010},
  editor = {Teh, Yee Whye and Titterington, Mike},
  volume = {9},
  series = {Proceedings of Machine Learning Research},
  publisher =    {PMLR},
}

@inproceedings{pytorch2019,
	author = {Paszke, Adam and Gross, Sam and Massa, Francisco and Lerer, Adam and Bradbury, James and Chanan, Gregory and Killeen, Trevor and Lin, Zeming and Gimelshein, Natalia and Antiga, Luca and Desmaison, Alban and Kopf, Andreas and Yang, Edward and DeVito, Zachary and Raison, Martin and Tejani, Alykhan and Chilamkurthy, Sasank and Steiner, Benoit and Fang, Lu and Bai, Junjie and Chintala, Soumith},
	booktitle = {Advances in Neural Information Processing Systems},
	publisher = {Curran Associates, Inc.},
	title = {PyTorch: An Imperative Style, High-Performance Deep Learning Library},
	volume = {32},
	year = {2019}
}

@inproceedings{kingma2015adam,
	author = {Kingma, Diederik P. and Ba, Jimmy},
	title = {Adam: A method for stochastic optimization},
	booktitle = {International Conference on Learning Representations (ICLR2015)},
        year = {2015},
	publisher = {Ithaca, NY},
	type = {Conference Proceedings}
}

@article{Jagtap2020,
	author = {Jagtap, Ameya D. and Kawaguchi, Kenji and Karniadakis, George Em},
	title = {Adaptive activation functions accelerate convergence in deep and physics-informed neural networks},
	journal = {J. Comput. Phys.},
	volume = {404},
	pages = {109136},
	ISSN = {0021-9991},
	year = {2020},
	doi = {10.1016/j.jcp.2019.109136},
	type = {Journal Article}
}

@article{xuUnderstandingOvercomingSpectral2025a,
	title = {On Understanding and Overcoming Spectral Biases of Deep Neural Network Learning Methods for Solving {{PDEs}}},
	author = {Xu, Zhi-Qin John and Zhang, Lulu and Cai, Wei},
	year = 2025,
	month = jun,
	journal = {J. Comput. Phys.},
	volume = {530},
	pages = {113905},
	issn = {0021-9991},
	doi = {10.1016/j.jcp.2025.113905},
	urldate = {2025-12-09},
	langid = {american},
}

@article{huangFrequencyadaptiveMultiscaleDeep2025,
	title = {Frequency-Adaptive Multi-Scale Deep Neural Networks},
	author = {Huang, Jizu and You, Rukang and Zhou, Tao},
	year = 2025,
	month = mar,
	journal = {Comput. Methods Appl. Mech. Eng.},
	volume = {437},
	pages = {117751},
	issn = {00457825},
	doi = {10.1016/j.cma.2025.117751},
	urldate = {2025-11-28},
	langid = {english},
}

@InProceedings{Rahaman2019,
  title = {On the Spectral Bias of Neural Networks},
  author = {Rahaman, Nasim and Baratin, Aristide and Arpit, Devansh and Draxler, Felix and Lin, Min and Hamprecht, Fred and Bengio, Yoshua and Courville, Aaron},
  booktitle = {Proceedings of the 36th International Conference on Machine Learning},
  pages = {5301-5310},
  year = 	 {2019},
  volume = {97},
  month = {09--15 Jun},
  publisher = {PMLR},
}

@article{Wang2021a,
	author = {Wang, Sifan and Teng, Yujun and Perdikaris, Paris},
	title = {Understanding and mitigating gradient flow pathologies in physics-informed neural networks},
	journal = {SIAM J. Sci. Comput.},
	volume = {43},
	number = {5},
	pages = {A3055-A3081},
	ISSN = {1064-8275},
	year = {2021},
	doi = {10.1137/20M1318043},
	type = {Journal Article}
}

@article{Wang2021b,
	author = {Wang, Sifan and Wang, Hanwen and Perdikaris, Paris},
	title = {On the eigenvector bias of Fourier feature networks: From regression to solving multi-scale PDEs with physics-informed neural networks},
	journal = {Comput. Methods Appl. Mech. Eng.},
	volume = {384},
	pages = {113938},
	ISSN = {0045-7825},
	year = {2021},
	doi = {10.1016/j.cma.2021.113938},
	type = {Journal Article}
}

@book{Gray1990,
	title={Chemical oscillations and instabilities: non-linear chemical kinetics},
	author={Gray, Peter and Scott, Stephen K},
	year={1990},
	publisher={Oxford University Press}
}

@misc{Chebfun2014,
	title={Chebfun guide},
	author={Driscoll, Tobin A and Hale, Nicholas and Trefethen, Lloyd N},
	year={2014},
	publisher={Pafnuty Publications, Oxford}
}

@article{Cox2002,
	title = {Exponential Time Differencing for Stiff Systems},
	journal = {J. Comput. Phys.},
	volume = {176},
	number = {2},
	pages = {430-455},
	year = {2002},
	issn = {0021-9991},
	doi = {10.1006/jcph.2002.6995},
	author = {S.M. Cox and P.C. Matthews},
}

@article{Chen2021,
	author = {Chen, Chun-Teh and Gu, Grace X.},
	title = {Learning hidden elasticity with deep neural networks},
	journal = {Proc. Natl. Acad. Sci. U. S. A.},
	volume = {118},
	number = {31},
	pages = {e2102721118},
	year = {2021},
	doi = {10.1073/pnas.2102721118},
	type = {Journal Article}
}

@article{Eckermann2014,
	title = {Computational modeling of creep-based fatigue as a means of selecting lead-free solder alloys},
	journal = {Microelectron. Reliab.},
	volume = {54},
	number = {6},
	pages = {1235-1242},
	year = {2014},
	issn = {0026-2714},
	doi = {10.1016/j.microrel.2014.02.017},
	author = {J. Eckermann and S. Mehmood and H.M. Davies and N.P. Lavery and S.G.R. Brown and J. Sienz and A. Jones and P. Sommerfeld},
}

@article{Ren2024,
	author = {Ren, Xiaodan and Lyu, Xianrui},
	title = {Mixed form based physics-informed neural networks for performance evaluation of two-phase random materials},
	journal = {Eng. Appl. Artif. Intell.},
	volume = {127},
	pages={107250},
	ISSN = {09521976},
	year = {2024},
	doi = {10.1016/j.engappai.2023.107250},
	type = {Journal Article}
}

@article{Liu2024,
	author = {Liu, Bokai and Wang, Yizheng and Rabczuk, Timon and Olofsson, Thomas and Lu, Weizhuo},
	title = {Multi-scale modeling in thermal conductivity of Polyurethane incorporated with Phase Change Materials using Physics-Informed Neural Networks},
	journal = {Renew. Energy},
	volume = {220},
	pages = {119565},
	ISSN = {0960-1481},
	year = {2024},
	doi = {10.1016/j.renene.2023.119565},
	type = {Journal Article}
}

@inproceedings{Iyer2019,
	author = {Iyer, Akshay and Zhang, Yichi and Prasad, Aditya and Tao, Siyu and Wang, Yixing and Schadler, Linda and Brinson, L. Catherine and Chen, Wei},
	title = {Data-Centric Mixed-Variable Bayesian Optimization for Materials Design},
	volume = {2A},
	booktitle = {International Design Engineering Technical Conferences and Computers and Information in Engineering Conference},
	pages = {V02AT03A066},
	year = {2019},
	doi = {10.1115/DETC2019-98222},
	month = {08},
}

@article{DIC2015,
    author = {Blaber, J. and Adair, B. and Antoniou, A.},
	title = {Ncorr: Open-Source 2D Digital Image Correlation Matlab Software},
	journal = {Exp. Mech.},
	volume = {55},
	number = {6},
	pages = {1105-1122},
	ISSN = {1741-2765},
	year = {2015},
	doi = {10.1007/s11340-015-0009-1},
	type = {Journal Article}
}

@book{Langtangen2016,
	author = {Langtangen, Hans Petter and Pedersen, Geir K.},
	title = {Scaling of differential equations},
	publisher = {Springer Nature},
	ISBN = {3319327267},
	year = {2016},
	doi = {10.1007/978-3-319-32726-6},
	type = {Book}
}

@article{FIAS2023,
	author = {Gao, Zhiwei and Yan, Liang and Zhou, Tao},
	title = {Failure-Informed Adaptive Sampling for PINNs},
	journal = {SIAM J. Sci. Comput.},
	volume = {45},
	number = {4},
	pages = {A1971-A1994},
	ISSN = {1064-8275},
	year = {2023},
	doi = {10.1137/22M1527763},
	type = {Journal Article}
}

@article{Mao2023,
	author = {Mao, Zhiping and Meng, Xuhui},
	title = {Physics-informed neural networks with residual/gradient-based adaptive sampling methods for solving partial differential equations with sharp solutions},
	journal = {Appl. Math. Mech.},
	volume = {44},
	number = {7},
	pages = {1069-1084},
	ISSN = {1573-2754},
	year = {2023},
	doi = {10.1007/s10483-023-2994-7},
	type = {Journal Article}
}

@article{DAS2023,
	author = {Tang, Kejun and Wan, Xiaoliang and Yang, Chao},
	title = {DAS-PINNs: A deep adaptive sampling method for solving high-dimensional partial differential equations},
	journal = {J. Comput. Phys.},
	volume = {476},
	pages = {111868},
	ISSN = {0021-9991},
	year = {2023},
	doi = {10.1016/j.jcp.2022.111868},
	type = {Journal Article}
}

@article{Wu2023,
	author = {Wu, Chenxi and Zhu, Min and Tan, Qinyang and Kartha, Yadhu and Lu, Lu},
	title = {A comprehensive study of non-adaptive and residual-based adaptive sampling for physics-informed neural networks},
	journal = {Comput. Methods Appl. Mech. Eng.},
	volume = {403},
	pages = {115671},
	ISSN = {0045-7825},
	year = {2023},
	doi = {10.1016/j.cma.2022.115671},
	type = {Journal Article}
}

@article{Rigas2024,
	title = {Adaptive {{Training}} of {{Grid-Dependent Physics-Informed Kolmogorov-Arnold Networks}}},
	author = {Rigas, Spyros and Papachristou, Michalis and Papadopoulos, Theofilos and Anagnostopoulos, Fotios and Alexandridis, Georgios},
	year = 2024,
	journal = {IEEE Access},
	volume = {12},
	pages = {176982--176998},
	issn = {2169-3536},
	doi = {10.1109/ACCESS.2024.3504962},
	urldate = {2025-03-05},
	langid = {american},
}

@article{Nabian2021,
	author = {Nabian, Mohammad Amin and Gladstone, Rini Jasmine and Meidani, Hadi},
	title = {Efficient training of physics-informed neural networks via importance sampling},
	journal = {Comput.-Aided Civ. Infrastruct. Eng.},
	volume = {36},
	number = {8},
	pages = {962-977},
	ISSN = {1093-9687},
	year = {2021},
	doi = {10.1111/mice.12685},
	type = {Journal Article}
}

@article{DeepXDE2021,
	author = {Lu, Lu and Meng, Xuhui and Mao, Zhiping and Karniadakis, George Em},
	title = {DeepXDE: A Deep Learning Library for Solving Differential Equations},
	journal = {SIAM Rev.},
	volume = {63},
	number = {1},
	pages = {208-228},
	ISSN = {0036-1445},
	year = {2021},
	doi = {10.1137/19M1274067},
	type = {Journal Article}
}

@article{Aldirany2024,
	author = {Aldirany, Ziad and Cottereau, Régis and Laforest, Marc and Prudhomme, Serge},
	title = {Multi-level neural networks for accurate solutions of boundary-value problems},
	journal = {Comput. Methods Appl. Mech. Eng.},
	volume = {419},
	pages = {116666},
	ISSN = {0045-7825},
	year = {2024},
	doi = {10.1016/j.cma.2023.116666},
	type = {Journal Article}
}

@article{Song2025,
	author = {Song, Suihong and Mukerji, Tapan and Zhang, Dongxiao},
	title = {Physics-informed multi-grid neural operator: Theory and an application to porous flow simulation},
	journal = {J. Comput. Phys.},
	volume = {520},
	pages = {113438},
	ISSN = {0021-9991},
	year = {2025},
	doi = {10.1016/j.jcp.2024.113438},
	type = {Journal Article}
}

@article{Wang2024,
	author = {Wang, Yongji and Lai, Ching-Yao},
	title = {Multi-stage neural networks: Function approximator of machine precision},
	journal = {J. Comput. Phys.},
	volume = {504},
	pages = {112865},
	ISSN = {0021-9991},
	year = {2024},
	doi = {10.1016/j.jcp.2024.112865},
	type = {Journal Article}
}

@article{wangArtificialIntelligencePartial,
  title = {Artificial {{Intelligence For Partial Differential Equations In Computational Mechanics}}: {{A Review}}},
  shorttitle = {Artificial {{Intelligence For Partial Differential Equations In Computational Mechanics}}},
  author = {Wang, Yizheng and Bai, Jinshuai and Lin, Zhongya and Wang, Qimin and Anitescu, Cosmin and Sun, Jia and Eshaghi, Mohammad Sadegh and Gu, YuanTong and Feng, Xiqiao and Zhuang, Xiaoying and Rabczuk, Timon and Liu, Yinghua},
  journal = {Appl. Mech. Rev.},
  pages = {1-81},
  doi = {10.1115/1.4071710},
  year = {2026},
  langid = {english},
}

@article{wangCENNConservativeEnergy2022,
  title = {{{CENN}}: {{Conservative}} Energy Method Based on Neural Networks with Subdomains for Solving Variational Problems Involving Heterogeneous and Complex Geometries},
  shorttitle = {{{CENN}}},
  author = {Wang, Yizheng and Sun, Jia and Li, Wei and Lu, Zaiyuan and Liu, Yinghua},
  year = 2022,
  month = oct,
  journal = {Comput. Methods Appl. Mech. Eng.},
  volume = {400},
  pages = {115491},
  issn = {0045-7825},
  doi = {10.1016/j.cma.2022.115491},
  urldate = {2025-12-22},
  langid = {american},
}

@article{wangDCEMDeepComplementary2024,
  title = {{{DCEM}}: {{A}} Deep Complementary Energy Method for Linear Elasticity},
  shorttitle = {{{DCEM}}},
  author = {Wang, Yizheng and Sun, Jia and Rabczuk, Timon and Liu, Yinghua},
  year = 2024,
  journal = {Int. J. Numer. Methods Eng.},
  volume = {125},
  number = {24},
  pages = {e7585},
  issn = {1097-0207},
  doi = {10.1002/nme.7585},
  urldate = {2026-05-23},
  copyright = {\copyright{} 2024 John Wiley \& Sons Ltd.},
  langid = {english},
}

@article{wangKolmogorovArnoldInformed2025,
	title = {Kolmogorov--Arnold-Informed Neural Network: A Physics-Informed Deep Learning Framework for Solving Forward and Inverse Problems Based on Kolmogorov--Arnold Networks},
	author = {Wang, Yizheng and Sun, Jia and Bai, Jinshuai and Anitescu, Cosmin and Eshaghi, Mohammad Sadegh and Zhuang, Xiaoying and Rabczuk, Timon and Liu, Yinghua},
	year = {2025},
	journal = {Comput. Methods Appl. Mech. Eng.},
	volume = {433},
	pages = {117518},
	doi = {10.1016/j.cma.2024.117518},
}

@article{le-ducNormalizedEnergybasedPhysicsinformed2026,
	title = {Normalized Energy-Based Physics-Informed Neural Network: {{Theory}} and Applications to Solid Mechanics Problems},
	shorttitle = {Normalized Energy-Based Physics-Informed Neural Network},
	author = {{Le-Duc}, Thang and {Nguyen-Xuan}, H. and Lee, Jaehong},
	year = 2026,
	month = apr,
	journal = {Finite Elem. Anal. Des.},
	volume = {256},
	pages = {104523},
	issn = {0168-874X},
	doi = {10.1016/j.finel.2026.104523},
	urldate = {2026-05-24},
}

@article{LingHu2025,
	title = {High-order multi-scale neural network method for quasi-static thermo-mechanical problems of composite materials},
	journal = {Appl. Math. Model.},
	volume = {148},
	pages = {116232},
	year = {2025},
	issn = {0307-904X},
	doi = {10.1016/j.apm.2025.116232},
	author = {Linghu,Jiale  and  Gao,Weifeng and Dong,Hao and Nie,Yufeng  and Wang, Shuqi},
}

@article{anagnostopoulosLearningPINNsPhase2026,
	title = {Learning in {{PINNs}}: {{Phase}} Transition, Diffusion Equilibrium, and Generalization},
	shorttitle = {Learning in {{PINNs}}},
	author = {Anagnostopoulos, Sokratis J. and Toscano, Juan Diego and Stergiopulos, Nikolaos and Karniadakis, George Em},
	year = 2026,
	month = jan,
	journal = {Neural Netw.},
	volume = {193},
	pages = {107983},
	issn = {0893-6080},
	doi = {10.1016/j.neunet.2025.107983},
	urldate = {2026-06-22},
}

@inproceedings{NEURIPS2018,
	title = {Neural Tangent Kernel: {{Convergence}} and Generalization in Neural Networks},
	booktitle = {Advances in Neural Information Processing Systems},
	author = {Jacot, Arthur and Gabriel, Franck and Hongler, Clement},
	editor = {Bengio, S. and Wallach, H. and Larochelle, H. and Grauman, K. and {Cesa-Bianchi}, N. and Garnett, R.},
	year = 2018,
	volume = {31},
	publisher = {Curran Associates, Inc.}
}

@article{wangWhenWhyPINNs2022,
	title = {When and Why {{PINNs}} Fail to Train: {{A}} Neural Tangent Kernel Perspective},
	shorttitle = {When and Why {{PINNs}} Fail to Train},
	author = {Wang, Sifan and Yu, Xinling and Perdikaris, Paris},
	year = 2022,
	month = jan,
	journal = {J. Comput. Phys.},
	volume = {449},
	pages = {110768},
	issn = {00219991},
	doi = {10.1016/j.jcp.2021.110768},
	urldate = {2025-02-03},
	langid = {english},
}





\end{document}